\documentclass[twocolumn,superscriptaddress]{revtex4}
\usepackage{color}
\usepackage{graphics,graphicx,epsfig}
\usepackage{epsf,epstopdf,wrapfig}
\usepackage{amssymb,amsfonts,amsmath}
\include{epsf}
\usepackage{ifthen}

\newcommand{\beqn}{\begin{eqnarray}}
\newcommand{\eeqn}{\end{eqnarray}}
\newcommand{\beq}{\begin{equation}}
\newcommand{\eeq}{\end{equation}}

\usepackage{ifthen}

\newcommand{\panelone}{
\begin{figure}
\begin{center}
\ifthenelse{\equal{\onlycaption}{true}}{}{
\noindent\includegraphics[width=.9\linewidth]{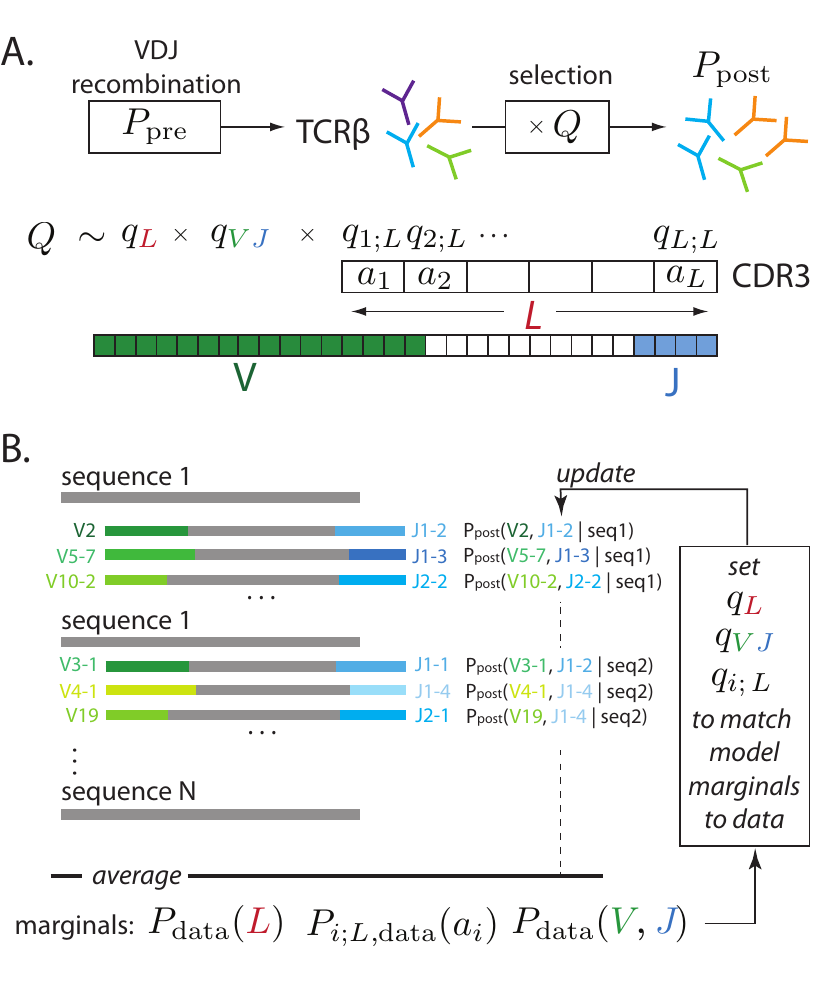}
}
\caption{{\bf A.} T-cell receptor $\beta$ chain sequences are formed during VDJ recombination. Sequences from this probability distribution, described by $P_{pre}$, are then selected with a factor $Q$ defined for each sequence, resulting in the observed $P_{post}$ distribution of receptor sequences. Selection is assumed to act independently on the V and J genes, the length of the CDR3 region and each of the amino acids, $a_i$, therein. {\bf B.} A schematic of the fitting procedure: the parameters are set so that $P_{\rm post}$ fits the marginal frequencies of amino acids at each position, the distribution of CDR3 lengths and VJ gene choices. Since the latter is not known unambiguously from the observed sequences, it is estimated probabilistically using the model itself in an iterative procedure.
\label{fig1}
}
\end{center}
\end{figure}
}

\newcommand{\paneltwo}{
\begin{figure}
\ifthenelse{\equal{\onlycaption}{true}}{}{
\noindent\includegraphics[width=\linewidth]{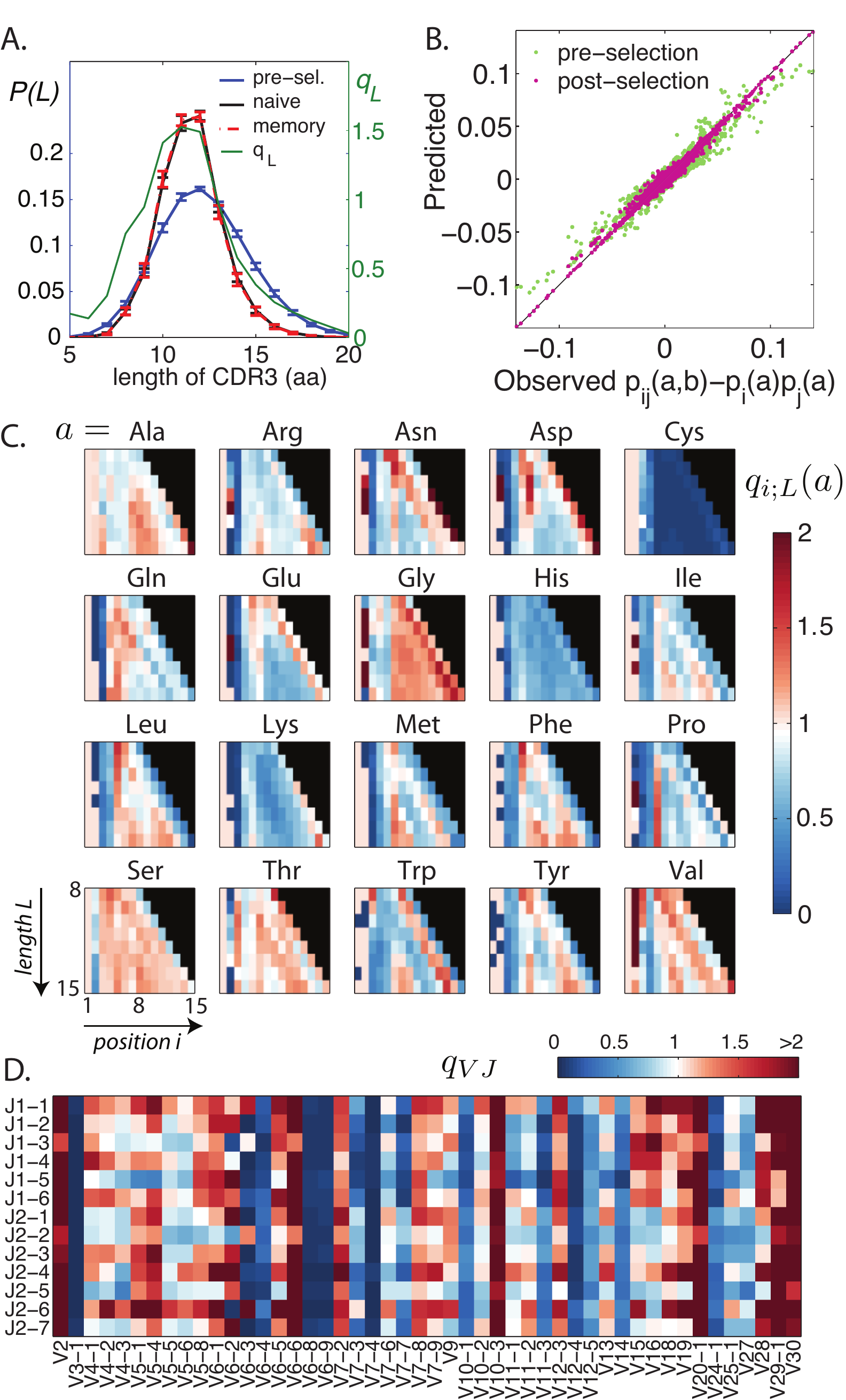}
}
\caption{
{\bf A.} CDR3 length distributions, pre- and post-selection and the length selection factor $q_L$ (green).
Selection makes the length-distribution of CDR3 regions in the pre-selection repertoire more peaked for the naive and memory repertoires (overlapping). Error bars show standard variation over $9$ individuals. {\bf B.} Comparison between data and model of the connected pairwise correlation functions, which were not fitted by our model. The excellent agreement validates the inference procedure. As a control, the prediction from the pre-selection model (in gray) does not agree with the data as well.
{\bf C.} Values of the inferred amino-acid selection factors for each amino acid, ordered by length of the CDR3 region (ordinate) and position in the region (abscissa). {\bf D.} Values of the $VJ$ gene selection factors.
\label{fig2}
}
\end{figure}
}

\newcommand{\panelthree}{
\begin{figure}
\ifthenelse{\equal{\onlycaption}{true}}{}{
\noindent\includegraphics[width=\linewidth]{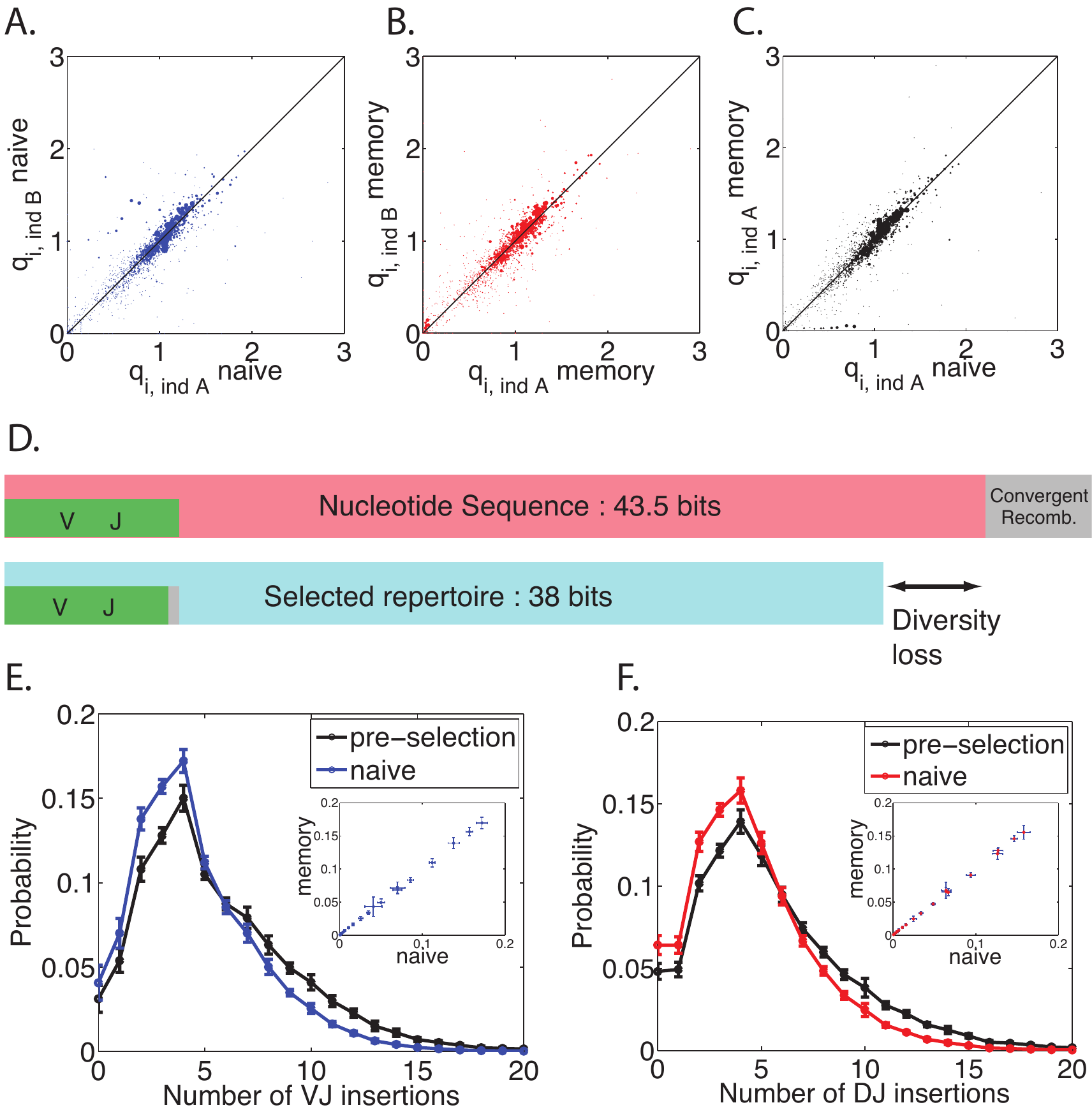}
}
\caption{{\bf A.-C.} Variability between repertoires. The scatter between $q_{i;L}$ selection factors between two sample individuals $A$ and $B$ for naive ({\bf A}) and memory repertoires ({\bf B}) compared to that of memory and naive repertoires for the same individual ({\bf C}) shows great similarity between them. See also Fig.~11.
{\bf D.} The entropy of the pre-selection repertoire (top) is reduced in the post-selection repertoire (bottom). 
{\bf E.-F.} Distribution of $VJ$ ({\bf E}) and $DJ$ ({\bf F}) insertions in the pre-selection and naive repertoires shows elimination of long insertions. Error bars show standard deviations over $9$ donors. The insertion distributions for the memory repertoire are the same as for the naive repertoire (see scatter plots in insets).
\label{fig3}
}
\end{figure}
}

\newcommand{\panelfour}{
\begin{figure}
\ifthenelse{\equal{\onlycaption}{true}}{}{
\noindent\includegraphics[width=\linewidth]{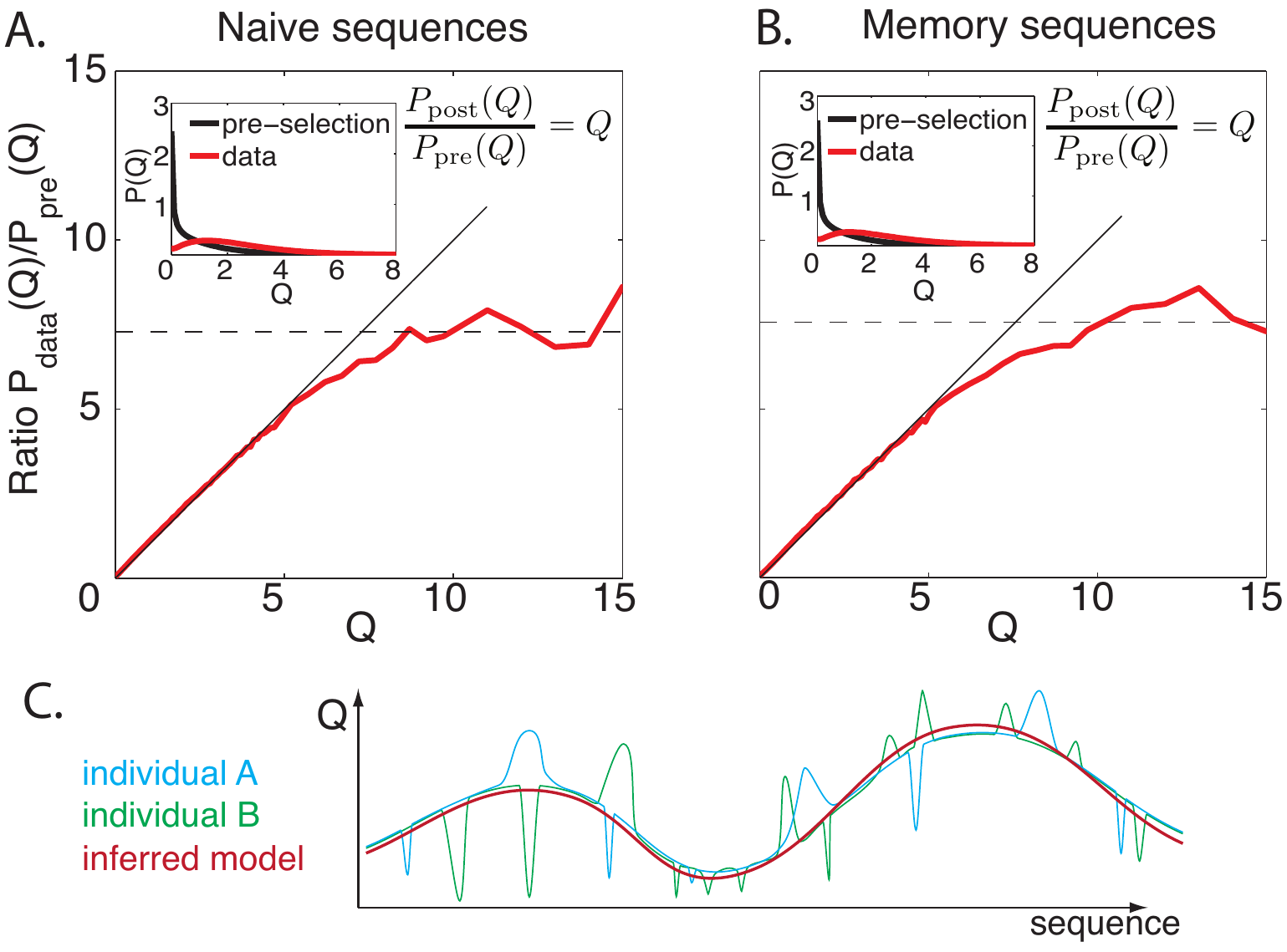}
}
\caption{
Probability of passing selection. {\bf A.-B.}  Ratio of the distributions of sequence-wide selection factors $Q$ between the observed sequences and the pre-selection ensemble (red line), plotted as a function of $Q$ for naive ({\bf A.}) and memory ({\bf B.}) repertoires. The model prediction $P_{\rm post}(Q)/P_{\rm pre}(Q)=Q$ is shown in black, and the pre-selection and observed distributions of Q are shown in the insets. The selection ratio saturate around $\approx 7$, which may be interpreted as the maximum probability of being selected. Naive and memory repertoires show similar behaviors.
{\bf C.} A cartoon of the effective selection landscape captured by our model (red line). Our method does not capture localized selection pressures (such as avoiding self) specific to each individual, but captures general global properties.
\label{fig4}
}
\end{figure}
}

\newcommand{\panelfive}{
\begin{figure}
\ifthenelse{\equal{\onlycaption}{true}}{}{
\noindent\includegraphics[width=\linewidth]{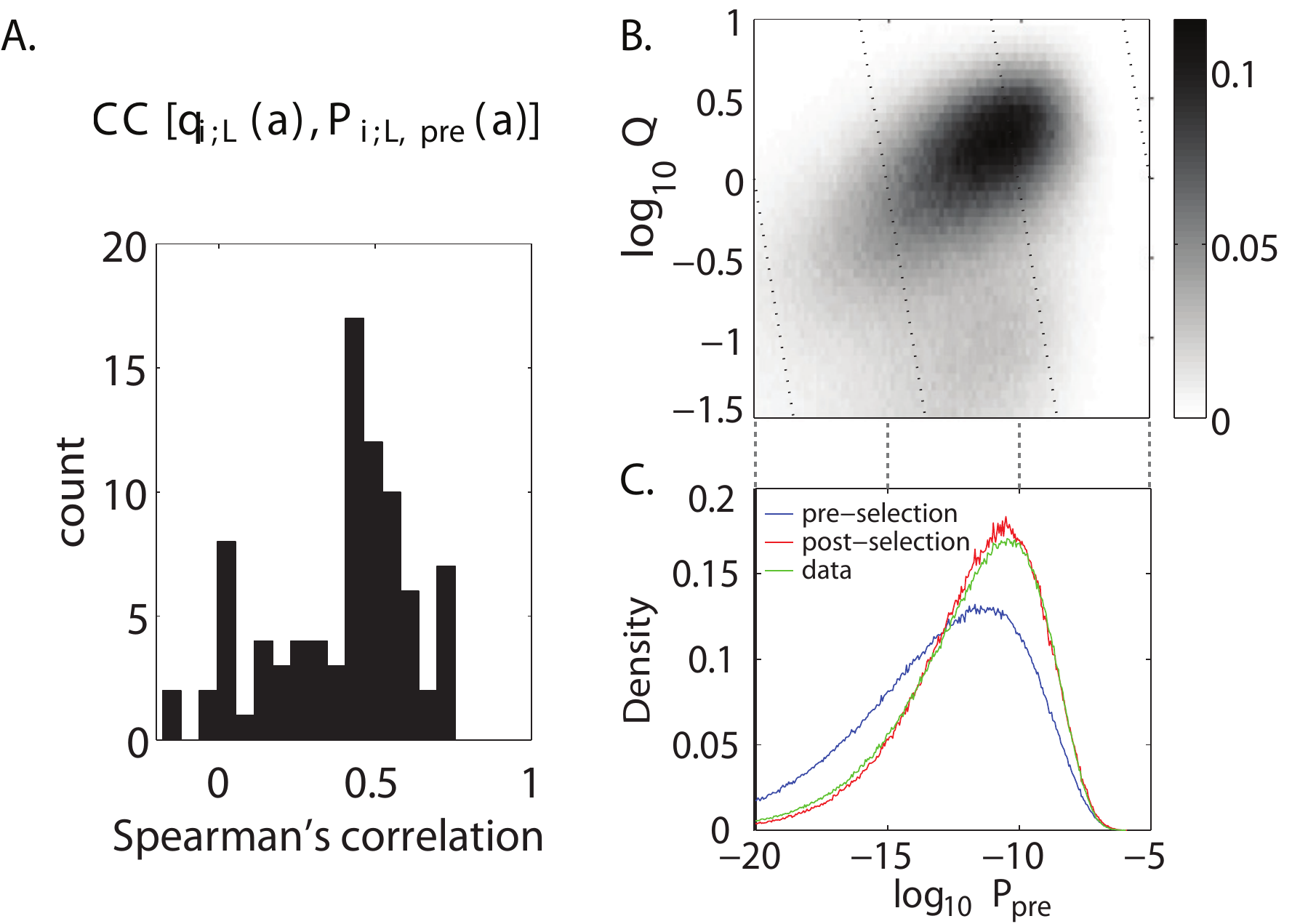}
}
\caption{
Correlations between the pre- and post-selection repertoires. {\bf A.} A histogram of Spearman's correlation coefficient values between the $q_{i;L}(a)$ selection factors in the CDR3 region and their generation probabilities $P_{i;L,pre}(a)$ for all $i,L$ shows an abundance of positive correlations.  {\bf B.} Heatmap of the joint distribution of the pre-selection probability distribution $P_{\rm pre}$ and selection factors $Q$ for each sequence shows the two quantities are correlated.
{\bf C.} Sequences in the observed, selected repertoire (green line) had a higher probability to have been generated by recombination than unselected sequences (blue line). Agreement between the post-selection model (red line) and data distribution (green line) is a validation of the model.
\label{fig5}
}
\end{figure}
}

\newcommand{\panelsix}{
\begin{figure}
\ifthenelse{\equal{\onlycaption}{true}}{}{
\noindent\includegraphics[width=\linewidth]{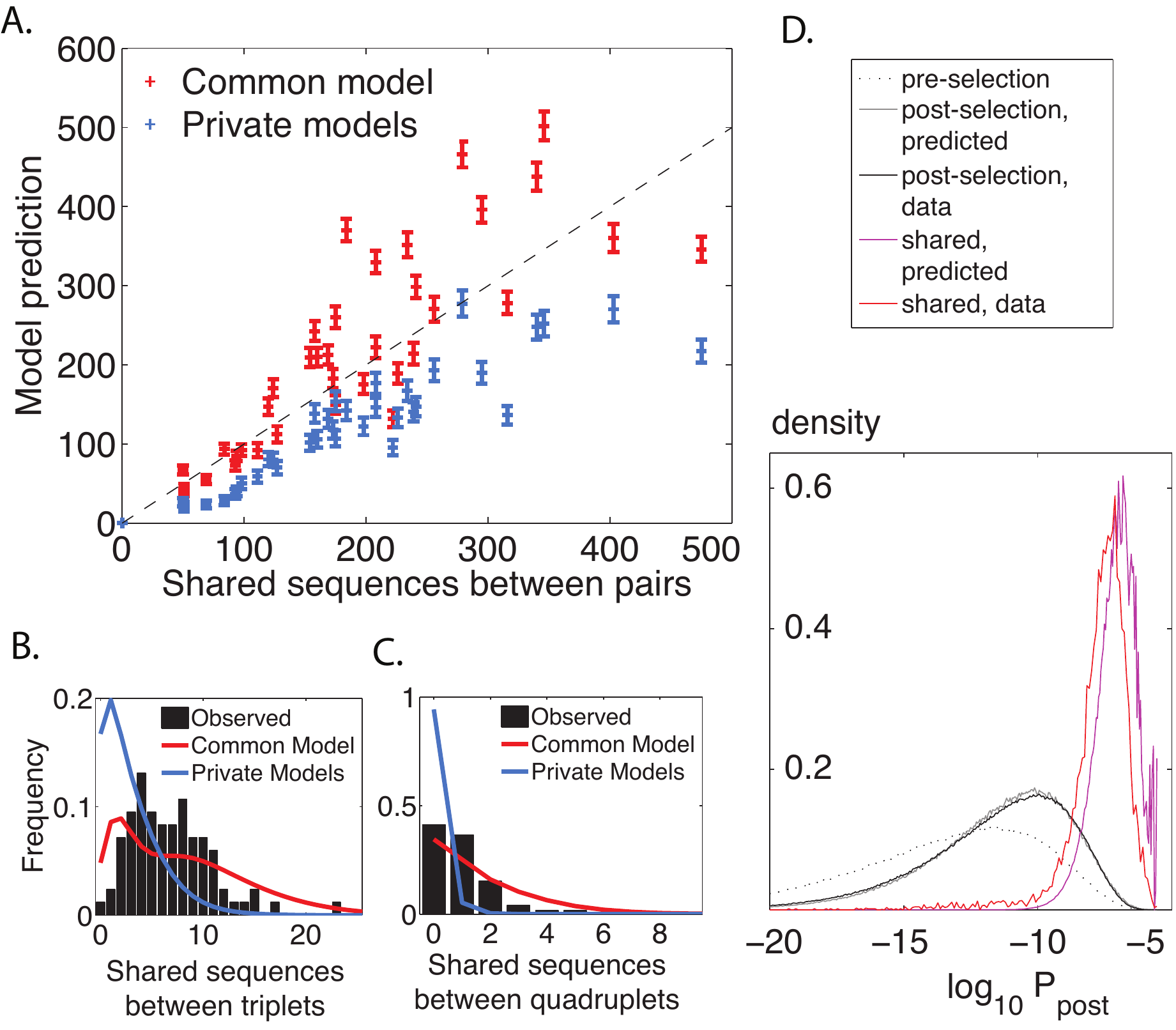}
}
\caption{
Shared sequences between individuals.  {\bf A.} The mean number of shared sequences between any pair of individuals compared to the number expected by chance (model prediction) for one common model for all individuals (red crosses) and private models learned independently for each individual (blue crosses). Error bars are standard deviations from distributions over pairs.  The distribution of shared sequences between triplets ({\bf B.})  and quadruplets ({\bf C.}) of individuals for the data (black histogram), from common (red line) and private models (blue line).   {\bf D.}  The shared sequences are most likely to be generated and selected: comparison of the $P_{post}$  post-selection distribution for sequences from the pre-selection (dotted line), and post-selection repertoires (according to the model in gray, and to the data in black), as well as the sequences shared by at least two donors (model prediction in magenta, data in red).
\label{fig6}
}
\end{figure}
}

\newcommand{\format}{pnasfigintext}  

\newcommand{\onlycaption}{false}

\begin{document}

\title{Quantifying selection in immune receptor repertoires}




\author{Yuval Elhanati}
\affiliation{Laboratoire de physique th\'eorique, UMR8549, CNRS and
  \'Ecole normale sup\'erieure, 24, rue Lhomond, 75005 Paris, France}
\author{Anand Murugan}
\affiliation{Department of Applied Physics, Stanford University,
  California 94305 USA}
\author{Curtis G. Callan Jr.}
\affiliation{Joseph Henry Laboratories,
Princeton University, Princeton, New Jersey 08544 USA}
\author{Thierry Mora}
\affiliation{Laboratoire de physique statistique, UMR8550, CNRS and
  \'Ecole normale sup\'erieure, 24, rue Lhomond, 75005 Paris, France}
\author{Aleksandra M. Walczak}
\affiliation{Laboratoire de physique th\'eorique, UMR8549, CNRS and
  \'Ecole normale sup\'erieure, 24, rue Lhomond, 75005 Paris, France}


\date{\today}

\begin{abstract}
The efficient recognition of pathogens by the adaptive immune system relies on the diversity of receptors displayed at the surface of immune cells. T-cell receptor diversity results from an initial random DNA editing process, called VDJ recombination, followed by functional selection of cells according to the interaction of their surface receptors with self and foreign antigenic peptides. To quantify the effect of selection on the highly variable elements of the receptor, we apply a probabilistic maximum likelihood approach to the analysis of high-throughput sequence data from the $\beta$-chain of human T-cell receptors. We quantify selection factors for V and J gene choice, and for the length  and amino-acid composition of the variable region. Our approach is necessary to disentangle the effects of selection from biases inherent in the recombination process. Inferred selection factors differ little between donors, or between naive and memory repertoires. The number of sequences shared between donors is well-predicted by the model, indicating a purely stochastic origin of such ``public'' sequences.
We find a significant correlation between biases induced by VDJ recombination and our inferred selection factors, together with a reduction of diversity during selection. Both effects suggest that natural selection acting on the recombination process has anticipated the selection pressures experienced during somatic evolution.
\end{abstract}

\maketitle

\noindent\fbox{
\begin{minipage}{.95\linewidth}
\vspace{0.2cm}
\section*{Significance statement}
The immune system defends against pathogens via a diverse population of T-cells that display different antigen recognition surface receptor proteins. Receptor diversity is produced by an initial random gene recombination process, followed by selection for a desirable range of peptide binding.  Although recombination is well-understood, selection has not been quantitatively characterized. By combining high throughput sequencing data with modeling, we quantify the selection pressure that shapes functional repertoires. Selection is found to vary little between individuals or between the naive and memory repertoires. It reinforces the biases of the recombination process, meaning that sequences more likely to be produced are also more likely to pass selection. The model accounts for ``public'' sequences shared between individuals as resulting from pure chance.
\vspace{0.2cm}
\end{minipage}
}

\bigskip


The T-cell response of the adaptive immune system begins when receptor proteins on the surface of 
these cells recognize a pathogen peptide presented by an antigen presenting cell. 
The immune cell repertoire of a given individual is comprised of 
many clones, each with a distinct surface receptor. This diversity, which is central to the ability of the immune system to defeat 
pathogens, is initially created 
by a stochastic process of germline DNA editing (called VDJ recombination) that gives each new immune cell a unique surface receptor gene. This initial repertoire is subsequently modified by selective forces, including thymic selection against excessive (or insufficient) recognition of self proteins, that are also stochastic in nature. Due to this stochasticity and the large T-cell diversity, these repertoires are best described by probability distributions. 
In this paper we apply a probabilistic approach to sequence data to 
obtain quantitative measures of the selection pressures that shape T-cell receptor repertoires.

New receptor genes are formed by choosing at random from a set of genomic templates for several sub-regions (V, D and J) of the complete gene. Insertion and deletion of nucleotides in the junctional regions between the V and D and D and J genes greatly enhances diversity beyond pure VDJ combinatorics \cite{Murugan02102012}. The variable region of the gene lies between the last amino acids of the V segment and the beginning of the J segment; it codes for the Complementarity Determining Region 3 (CDR3) loop of the receptor protein, a region known to be functionally important in recognition \cite{Janeway:2005te}. Previous studies have shown that immune cell receptors are not uniform in terms of VDJ gene segment usage \cite{Weinstein:2009p1566, Ndifon:2012p12787, Mora:2010jx, Quigley:2010p12636}, or probability of generation \cite{Murugan02102012}, and that certain receptors are more likely than others to be shared by different individuals \cite{Venturi:2008p13151,Ndifon:2012p12787}. In other words, the statistical properties of the immune repertoire are rather complex, and their accurate determination requires sophisticated methods.


Over the past few years, advances in sequencing technology have made it possible to sample the T-cell receptor diversity of individual subjects in great depth \cite{Baum:2012p13186}, and this has in turn led to the development of sequence statistics-based approaches to the study of immune cell diversity \cite{Six:2013p13184,Robins:2013p13119}. In particular, we recently quantitatively characterized the primary, pre-selection diversity of the human T-cell repertoire by learning the probabilistic rules of VDJ recombination from out-of-frame DNA sequences that cannot be subject to functional selection and whose statistics can reflect only the recombination process \cite{Murugan02102012}.  After generation, T-cells undergo a somatic selection process in the thymus \cite{Yates:2014p13194} and later in the periphery \cite{Jameson:2002p13189}. Cells that pass thymic selection enter the peripheral repertoire as `naive' T-cells, and the subset of naive cells that eventually engage in an immune response will survive as a long-lived \lq memory\rq\, pool. Even though we now understand the statistical properties of the initial repertoire of immune receptors \cite{Murugan02102012}, and despite some theoretical studies of thymic selection at the molecular level \cite{Detours:1999p13187,Kosmrlj:2008p5340},
a quantitative understanding of how selection modifies those statistics to produce the naive and memory repertoires is lacking. 

In this paper, we build on our understanding of the primitive pre-selection distribution of T-cell receptors to derive a statistical method for identifying and quantifying selection pressures in the adaptive immune system. We apply this method to naive and memory DNA sequences of human T-cell $\beta$ chains obtained from peripheral blood samples of nine healthy individuals (Fig.\,\ref{fig1}). Our goal is to characterize the likelihood that any given sequence, once generated, will survive selection. 
Our analysis reveals strong and reproducible signatures of selection on specific amino acids in the CDR3 sequence, and on the usage of V and J genes. Most strikingly, we find significant correlation between the primitive generation probability of a sequence and the probability it will pass selection. This suggests that natural selection, which acts on very long time scales to shape the generation mechanism itself, may have tuned it to anticipate somatic selection, which acts on single cells throughout the lifetime of an individual. 
The quantitative features of selection inferred from our model vary very little between donors, indicating that these features are universal. In addition, our measures of selection pressure on the memory and naive repertoires are statistically indistinguishable, consistent with the hypothesis that the memory pool is a random subsample of the naive pool.

\section{Methods}

We analyzed human CD4+ T-cell  $\beta$-chain DNA sequence reads (60 or 101 nucleotides long) centered around the CDR3 region. T-cells were obtained from nine individuals and sorted into naive (CD45RO-) and memory (CD45RO+) subsets, yielding datasets of $\sim$200,000 unique naive and $\sim$120,000 unique memory sequences per individual, on average. The datasets are the same as those used in \cite{Murugan02102012} and were obtained by previously described methods \cite{Robins:2009da,Robins:2010hd}. 

In \cite{Murugan02102012} we used the ``nonproductive'' sequences (those where either the J sequences are out of frame, or the CDR3 sequences have a stop codon) to characterize the receptor generation process. The result of that analysis was an evaluation of the probability $P_{\rm pre}(\vec{\sigma})$ that a VDJ recombination event will produce a $\beta$-chain gene consistent with the specific DNA sequence read $\vec{\sigma}$. In this study we focus instead on the in-frame, productive, sequences (from both the naive and the memory repertoires) with the goal of quantifying how the post-selection probability distribution on sequences is modified from the original distribution $P_{\rm pre}(\vec{\sigma})$. In what follows we distinguish between the read $\vec\sigma$ and the CDR3 region $\vec\tau$, the latter defined to run from a conserved cysteine near the end of the V segment to the last amino acid of the read (leaving two amino acids to the conserved Phe). The CDR3 amino acid sequence can be uniquely read off from each in-frame sequence read; by contrast, the specific V- and J-genes responsible for the read may not be uniquely identifiable (because of the relatively short read length). An unambiguous selection effect can be seen by comparing the length distribution of CDR3 regions between the pre-selection ensemble and the naive, or memory, datasets (Fig.\,\ref{fig2}A): sequences that are longer or shorter than the mean are suppressed resulting in a more peaked distribution. 

For each receptor sequence, we define a selection factor $Q(\vec\sigma)$ that quantifies whether selection (thymic selection or later selection in the periphery) has enriched or impoverished the frequency of $\vec\sigma$ compared to the pre-selection ensemble. Since the generation probability of sequences varies over many orders of magnitude, such a comparison is the only way to define selection strength. Denoting by $P_{\rm post}(\vec\sigma)$ the distribution of sequences in the selected naive or memory pools, we will set $P_{\rm post}(\vec\sigma)=Q(\vec\sigma)P_{\rm pre}(\vec\sigma)$. Due to the large number of possible sequences, we cannot sample the post-selection probability $P_{\rm post}$ for each sequence directly from the data; we need a reduced complexity model to estimate it. We propose a simple model, summarized in Fig.\,\ref{fig1}A, that we we will show captures the main features of selection:
\beq
Q(\vec{\tau},V,J)=\frac{P_{\rm post}(\vec{\tau},V,J)}{P_{\rm pre}(\vec{\tau},V,J)}=
\frac{1}{Z} \ q_L \   q_{VJ} \ \prod_{i=1}^L q_{i;L}(a_i),
\label{Qdef}
\eeq
where $V$ and $J$ denote the choice of $V$ and $J$ segments in the sequence $\vec\sigma$, $L$ is the amino-acid length of the CDR3  specified by the read, $(\tau_1,\ldots,\tau_{3L})$ is CDR3 nucleotide sequence, and $(a_1,\ldots,a_L)$ its amino-acid sequence.
The factors $q_L$, $q_{i;L}(a)$ and $q_{VJ}$  denote, respectively, selective pressures on the CDR3 length, its composition, and the associated VJ identities. Note that the D segment is entirely included in this junctional region, so selection acting on it is encoded in the $q_{i;L}$ factors. $Z$ enforces the model normalization condition $\sum_{\vec{\tau},V,J} Q(\vec{\tau},V,J) P_{\rm pre}(\vec{\tau},V,J)=1$.

\ifthenelse{\equal{\format}{pnasfigend}}{}{\panelone}

It is important to understand why we do not write $Q$ directly as a function of the read $\vec\sigma$. While $(\vec\tau,V,J)$ determines $\vec\sigma$ and $\vec\sigma$ determines $\vec\tau$, $V$ and $J$ cannot always be inferred deterministically from the read $\vec\sigma$. The VJ assignment of any given read will have to be treated as probabilistically defined hidden variables. In addition, because of correlations in  $P_{\rm pre}$, the $q$ factors cannot be identified with marginal enrichment factors (so that, for example, $P_{\rm i;L,\rm data} (a_i)/ P_{i;L,\rm pre} (a_i)$, cannot be set equal to $q_{i;L}(a_i)$). For all these reasons, we must use a maximum likelihood procedure to learn the $q_L$, $q_{i;L}$ and  $q_{V,J}$ factors of Eq.\,\ref{Qdef}. We use an expectation maximization algorithm (EM) that iteratively modifies the $q's$ until the observed marginal frequencies (for CDR3 length, amino acid usage as a function of CDR3 position, and VJ usage) in the data match those implied by the model distribution Eq.\,\ref{Qdef} (the pre-selection distribution $P_{\rm pre}$ being taken as a fixed, known, input). The procedure is schematically depicted in Fig.\,\ref{fig1}B (see the appendices for full details). 

One important assumption of the model is that selection factors act independently of each other on the sequence. Consequently, while the model is fit only to single point marginal frequencies, and not to pairwise frequencies. To check the validity of this assumption, we plot the correlation functions of amino acid pairs in the model post-selection repertoire vs the observed naive ones (Fig\,\ref{fig2}B). These pairwise correlations are well predicted, even though they are not model inputs. It is also noteworthy that they are nonzero, even though the selection model does not take into account the possibility of interactions in the selection factors $q_{i;L}$. This is because the pre-selection distribution does not factorize over amino acids in the CDR3 region, and has correlations of its own, as shown by the green points of Fig~\ref{fig2}B (note that these pre-selection correlations do not agree well with those observed in the post-selection data). 

Another assumption of our model is that selection acts at the level of the amino acid sequence, regardless of the underlying codons. To test this, we learned more general models where $a$ represented one of the possible 61 codons, instead of one of the 20 amino acids. We found that codons coding for the same residue had similar selection factors (see Fig.~8), except near the edges of the CDR3 where amino acids may actually come from genomic V and J segments and reflect their codon biases.

To compare the different donors, we learned a distinct model for each donor, as well as a ``universal'' model for all sequences of a given type from all donors taken together (see the appendices for details). We also learned models from random subsets of the sequence dataset to assess the effects of low-number statistical noise.

\ifthenelse{\equal{\format}{pnasfigend}}{}{\paneltwo}

\section{Results}

\subsection{Characteristics of selection and repertoire variability}

The length, single-residue, and VJ selection factors, learned from the naive datasets of all donors taken together, are presented in Fig.\,\ref{fig2}A,C,D. The $q_{VJ }$ distribution shows that the different $V$ and $J$ genes are subject to a wide range of selection factors (note that these factors act in addition to the quite varied gene segment usage probabilities in $P_{pre}(\vec\sigma)$).
We looked for correlations between the selection factors $q_{i;L}(a)$ on amino acids and a variety of amino-acid biochemical properties \cite{Stryer}: hydrophobicity, charge, pH, polarity, volume, and propensities to be found in $\alpha$ or $\beta$ structures, in turns, at the surface of a binding interface, on the rim or in the core \cite{MartinLavery} (see the appendices for details and references). We found no significant correlations, save for a negative correlation with amino acid volume and $\alpha$ helix association, as well as a positive correlation with the propensities to be in turns or in the core of an interacting complex (Fig.~13).

To estimate differences between datasets, we calculated the correlation coefficients between the logs of the $q_{VJ}$ and $q_{i;L}(a)$ selection factors (see Fig.~10). Comparing naive {\em vs.} naive, memory {\em vs.} memory or naive {\em vs.} memory between donors (see Fig.\,\ref{fig3}A-C for an example for $q_{i;L}$, and Fig.~9 for $q_{VJ}$) gave correlation coefficients of $\approx 0.9$ in $\log q_{i;L}$, while the naive {\em vs.} memory repertoires of the same donor gave $0.95$. To get a lower bound on small-number statistical noise, we also compared the factors inferred from artificial datasets obtained by randomly shuffling sequences between donors (see the appendices), yielding an average correlation coefficient of $0.98$. Repeating the analysis for $\log q_{VJ}$, we found correlation coefficients of $\approx 0.8$ between datasets of different donors, $0.84$ for the naive and memory dataset of the same donor, all of which must be compared to $0.94$ obtained between shuffled datasets. Thus, the observed variability between donors of $q_{i;L}$ and $q_{VJ}$ are small, and consistent with their expected statistical variability.

We use Shannon entropy,  $S= -\sum_{\vec{\sigma}} P_{\rm post}(\vec\sigma) \log_2{P_{\rm post}(\vec\sigma)}$, to quantify the diversity of the naive and memory distributions. Entropy is a diversity measure that accounts for non-uniformity of the distribution and is additive in independent components. Since $S=\log_2 \Omega$ when there are $\Omega$ equally likely outcomes, the diversity index $2^S$ can be viewed as an effective number of states. The entropy of the naive repertoire according to the model is $38\, {\rm bits}$ (corresponding to a diversity of $\sim 3.0 \cdot 10^{11}$) down from $43.5 \,{\rm bits}$ in the primitive, pre-selection repertoire (Fig.\,\ref{fig3}D). This is a reduction of $\sim 6\,{\rm bits}$, or $50$-fold in diversity. The majority of the reduction comes from insertions and deletions, which accounted for most of the diversity in the pre-selection repertoire. The entropies of the memory and naive repertoires are the same, indicating that selection in the periphery does not further reduce diversity.

\ifthenelse{\equal{\format}{pnasfigend}}{}{\panelthree}

Knowing the post-selection distribution of sequences, we can ask how different features of the recombination scenario fare in the face of selection. This does not imply that selection acts on the scenarios themselves ---it acts on the final product--- but it is an {\em a posteriori} assessment of the fitness of particular rearrangements. For example, the distributions of insertions at VD and DJ junctions in the post-selection ensemble have shorter tails (Fig.\,\ref{fig3}E-F), while the distribution of deletions at the junctions seems little affected by selection (Fig.~11), although large numbers of deletions are selected against.

\ifthenelse{\equal{\format}{pnasfigend}}{}{\panelfour}

\subsection{Selection factor as a measure of fitness}

The selection factor $Q$ is a proxy for the probability of a particular sequence to be selected or amplified, and sequences with large $Q$ values should thus be enriched in the observed dataset. To test this, we consider the distributions of $Q$ both in the pre-selection model, $P_{\rm pre}(Q)$, and in the dataset from which $Q$ was learned, $P_{\rm data}(Q)$ (insets of Fig.\,\ref{fig4}; see the appendices for details on how the distributions are calculated when V and J are hidden variables). This approach is very similar to the one used by Mustonen et al. \cite{Mustonen:2005p12433,Mustonen:2008p4315} to characterize the fitness landscape of transcription factor binding sites. 

By construction, the distribution of $Q$ in the post-selection model satisfies exactly $P_{\rm post}(Q)=QP_{\rm pre}(Q)$. 
In Fig.~\ref{fig4}A-B we plot the ratio $P_{\rm data}(Q)/P_{\rm pre}(Q)$ as a function of $Q$, both for the naive and memory models learned from all donors.
We observe that for $Q\leq 5$, {\em i.e.} for $>90\%$ of sequences, this ratio is exactly equal to $Q$ ---~a validation of our model prediction at the sequence-wide level. For larger values of $Q$ however, this ratio saturates to around $Q_{\rm max}\approx 7$.

This plateau may be viewed as a limiting value, above which selection is insensitive to $Q$. A similar plateau was observed in the fitness of transcription factor binding sites below a certain binding energy \cite{Mustonen:2008p4315}. In the case considered here, the plateau can be rationalized if we assume that $Q$ is proportional to the probability for a sequence to be selected, $P_{\rm sel}(\vec\sigma)=\alpha Q(\vec\sigma)$. Since $P_{\rm sel}$ cannot exceed one, $Q$ cannot exceed $\alpha^{-1}$. The average probability of selection is given by $\sum_{\vec\sigma}P_{\rm pre}(\vec\sigma)P_{\rm sel}(\vec\sigma)=\alpha$. The observed plateau gives a lower bound to the true maximum of $Q$: $\alpha^{-1}\geq Q_{\rm max}$, and thus the average fraction of cells to pass selection satisfies $\alpha\leq 15\%$. This can be compared to estimates \cite{Janeway:2005te} for passing positive and negative thymic selection: $10-30\%$ for positive selection only, and $\approx 5\%$ for both. This analysis only includes the $\beta$ chain, and including the $\alpha$ chain could further reduce our estimate.

The saturation also seems to indicate that our model may be too simple to describe the very fit (high $Q$) sequences. Because of its fairly simple factorized structure, our model can only account for the coarse features of selection, and may not capture very individual-specific traits such as avoidance of the self (corresponding to $Q\ll 1$ in localized regions of the sequence space) or response to pathogens ($Q\gg 1$ for particular sequences). This individual-dependent ruggedness of the fitness landscape $Q$, schematized in Fig.\,\ref{fig4}C, is probably ignored by our description, and may be hard to model in general.

To check that the saturation does not affect our inference procedure, we relearned our model parameters from simulated data, where sequences were generated from $P_{\rm pre}$ and then selected with probability $\min(Q/Q_{\rm max},1)$ (see the appendices for details), and we found that the model was correctly recovered (Fig.\,12).

\ifthenelse{\equal{\format}{pnasfigend}}{}{\panelfive}

\subsection{Natural selection anticipates somatic selection}

Comparing the pre- and post-selection length distributions in Fig.\,\ref{fig2}A shows that the CDR3 lengths that were the most probable to be produced by recombination are also more likely to be selected.
Formally, Spearman's rank correlation coefficient between $P_{\rm pre}(L)$ and $q_L$ is $0.76$, showing good correlation between the probability of a CDR3 length and the corresponding selection factor. 
We asked whether this correlation was also present in the other sequence features.
The histogram of Spearman's correlation between the selection factors $q_{i;L} (a)$ and the pre-selection amino-acid usage $P_{i;L,\rm pre}(a)$ for different lengths and positions $(i,L)$ (Fig.\,\ref{fig5}A) shows a clear majority of positive correlations. Likewise, the selection factors $q_{VJ}$ are positively correlated with the pre-selection VJ usage $P_{VJ,\rm pre}$ (Spearman's rank correlation $0.3$, $p<2\cdot 10^{-20}$).

The correlations observed for each particular feature of the sequence (CDR3 length, amino acid composition and VJ usage) combine to create a global correlation between the probability $P_{\rm pre}(\vec\sigma)$ that a sequence $\vec\sigma$ was generated by recombination, and its propensity $Q(\vec\sigma)$ to be selected (Spearman's rank correlation $0.4$, $p=0$, see Fig.\,\ref{fig5}B).
Consistent with this observation, the post-selection repertoire is enriched in sequences that have a high probability $P_{\rm pre}(\vec\sigma)$ to be produced by recombination (Fig.\,\ref{fig5}C). This enrichment is well predicted by the model, providing another validation of its predictions at the sequence-wide level.

Taken together, these results suggest that the mechanism of VDJ recombination (including insertions and deletions) has evolved to preferentially produce sequences that are more likely to be selected by thymic or peripheral selection.

\ifthenelse{\equal{\format}{pnasfigend}}{}{\panelsix}

\subsection{Shared sequences between individuals}
The observation of unique sequences that are shared between different donors has suggested that these sequences make up a ``public'' repertoire common to many individuals, formed through convergent evolution or a common source. However, it is also possible that these common sequences are just statistically more frequent, and are likely to be randomly recombined in two individuals independently, as previously discussed by Venturi {\em et al.} \cite{Venturi:2006p12655,Venturi:2008p13151,Quigley:2010p12636}. In other words, public sequences could just be chance events. Here we revisit this question by asking whether the number of observed shared sequences between individuals is consistent with 
random choice from our inferred sequence distribution $P_{\rm post}$.

We estimated the expected number of shared sequences between groups of donors in two ways: (i) by assuming that each donor $\alpha$ had its own ``private'' model learned from his own sequences or (ii) by assuming that sequences are drawn from a ``universal'' model learned from all sequences together. While the latter ignores small yet perhaps significant differences between the donors, the former may exaggerate them where statistics are poor.
For details on how these estimates are obtained from the models, we refer the reader to the appendices. In Fig.\,\ref{fig6}A we plot, for each pair of donors, the expected number of shared nucleotide sequences in their naive repertoires under assumptions (i) and (ii), versus the observed number. The number is well predicted under both assumptions, the universal model assumption giving a slight overestimate, and the private model giving a slight underestimate. We repeat the analysis for sequences that are observed to be common to at least three or at least four donors (Fig.\,\ref{fig6}B-C). The universal model predicts their number better than the private models, although it still slightly overestimates it.

These results suggest that shared sequences are indeed the result of pure chance. If that is so, shared sequences should have a higher occurrence probability than average; specifically, the model predicts that the sequences that are shared between at least two donors are distributed according to $P_{\rm post}(\vec\sigma)^2$ (see the appendices). We test this by plotting the distribution of $P_{\rm post}$ for regular sequences, as well as for pairwise-shared sequences, according to the model and in the naive datasets (Fig.\,\ref{fig6}D), and find excellent agreement. In general, sequences that are shared between at least $n$ individuals by chance should be distributed according to $ P_{\rm post}(\vec\sigma)^n$. For triplets and quadruplets, this model prediction is not as well verified (see Fig.\,14). This discrepancy may be explained by the fact that such sequences are outliers with very high occurrence probabilities, and may not be well captured by the model, which was learned on typical sequences.

We repeated these analyses for sequences shared between the memory repertoires of different individuals, with very similar conclusions, except for donors 2 and 3, and donors 2 and 7, who shared many more sequences than expected by chance (see Fig.\,15). We conclude that the vast majority of shared sequences occur by chance, and are well predicted by our model of random recombination and selection.

\section{Discussion}

We have introduced and calculated a selection factor $Q(\vec\sigma)$ that serves as a measure of selection acting on a given receptor sequence $\vec\sigma$ in the somatic evolution of the immune repertoire. Our approach accounts for the fact that the pre-selection probabilities of sequences vary over orders of magnitude.

Using this measure, we show that the observed repertoires have undergone significant selection starting from the initial repertoire produced by VDJ recombination. We find little difference between the naive and memory repertoires, in agreement with recent findings showing no correlation between receptor and T-cell fate \cite{Wang:2010p13192}, as well as between the repertoires of different donors. This is perhaps surprising, because the donors have distinct HLA types (which determine the interaction between T-cell receptors and peptide-MHC complexes), and we could expect their positive and negative selective pressures to be markedly different. Besides, memory sequences have undergone an additional layer of selection compared to the naive ones ---recognizing a pathogen--- and we could also expect to see different signatures of selection there. A possible interpretation is that our model only captures coarse and universal features of selection related to the general fitness of receptors, and not the fine-grained, individual-specific selective pressures such as avoidance of the self, as illustrated schematically in Fig.\,\ref{fig4}C. In other words, our selection factors may ``smooth out'' the complex landscapes of specific repertoires and fail to capture their rough local properties, such as would be expected from specific epitopes that would correspond to very tall peaks or deep valleys in the landscape of selection factors. To really probe these specific deep valleys, we need to develop methods based on accurate sequence counts. Another interesting future direction would be to see whether at this global level the signatures of selection are similar between (relatively) isolated populations. Lastly, comparing data from different species (mice, fish), in particular where inbred individuals with the same HLA type can be compared, would be an interesting avenue for addressing these issues. 

Our results suggest that natural selection has refined the generation process over evolutionary time scales to produce a pre-selection repertoire that anticipates the actions of selection. Sequences that are likely to be eliminated and fail selection are not very likely to be produced in the first place. Because of this ``rich become richer'' effect, the diversity of the repertoire is significantly reduced by selection, by a 50-fold factor in terms of diversity index. This does not mean that only 2\% of the sequences pass selection. In fact, our results are consistent with as much as $15\%$ of sequences passing selection. This apparent paradox is resolved by noting that selection, by keeping clones that were likely to be generated, get rids of very rare clones that contributed to the large initial diversity.

Although we did observe sequences that were present in the repertoires of different donors, we showed using our model that their number was broadly compatible with that expected by pure chance. This suggests that the ``public'' part of the repertoire is made of sequences that are just more likely to be randomly produced and selected.

To summarize, our work clearly shows that thymic selection and later peripheral selection modify the form of the  generated repertoire. Our work is a starting point for a description of a mechanism of the two processes.

{\bf Acknowledgements.} The work of YE, TM and AW was supported in part by grant ERCStG n. 306312. The work of CC was supported in part by NSF grants PHY-0957573 and PHY-1305525 and by W.M. Keck Foundation Award dated 12/15/09.

{\bf Acknowledgements.} The work of YE, TM and AW was supported in part by grant ERCStG n. 306312. The work of CC was supported in part by NSF grants PHY-0957573 and PHY-1305525 and by W.M. Keck Foundation Award dated 12/15/09.

\appendix

\section{Data}
The DNA nucleotide data used in our analysis consists of human CD4+ naive (CD45RO-) or memory (CD45RO+) $\beta$ chain sequences from $9$ healthy individuals, sequenced and made available to us by H. Robins and already used in \cite{Murugan02102012}. Reads are $60$ base pair long for $6$ donors and $101$ base pair long for $3$ donors (individuals $2, 3$ and $7$) and contain the CDR3 region and neighboring $V$ and $J$ gene nucleotides. All end at the same position in the $J$ gene, with four nucleotides between this position and the first nucleotide of the conserved phenylalanine. The data were divided into out-of-frame reads (non-coding), used to learn the pre-selection model as described in \cite{Murugan02102012} and in-frame (coding) reads used in the analysis presented in this paper. The sequence data we used are available at {\tt http://princeton.edu/\~{}ccallan/TCRPaper/data/}.

In our study we limit ourselves to unique sequences. The experimental procedure and initial assessment of the quality of the reads were done in the Robins lab following the procedures described in \cite{Robins:2010hda, Robins:2009da}. Each sequence was read multiple times, allowing for the correction of most sequencing errors. The numbers of unique sequences used in each dataset is shown in Table S\ref{table}.

\begin{table}[h!]
\begin{tabular}{|l||c|c|}
\hline
& Naive & Memory \\
\hline
\hline
Donor 1 & 311917 & 177744 \\
\hline
Donor 2 & 242254 & 135567 \\
\hline
Donor 3 & 195007 & 119906 \\
\hline
Donor 4 & 130958 & 142017 \\
\hline
Donor 5 & 147848 & 32468 \\
\hline
Donor 6 & 187245 & 104119 \\
\hline
Donor 7 & 251335 & 136419 \\
\hline
Donor 8 & 42326 & 120527 \\
\hline
Donor 9 & 254349 & 89830 \\
\hline
\end{tabular}
\caption{Number of unique coding sequences in each datasets.
\label{table}}
\end{table}

The alignment to all possible $V$ and $J$ genes was done using the curated datasets in the IMGT database \cite{Monod:2004im}. There are $48$ V genes, $2$ D genes and $13$ J genes plus a number of pseudo V genes that cannot lead to a functioning receptor due to stop codons. We discarded sequences that were associated to a pseudo-gene as our model only accounts for coding genes. The germline sequences of the genes used in our analysis are the same as were used in \cite{Murugan02102012} to analyze the generative V(D)J recombination process. The complete list of gene sequences can be found at {\tt http://princeton.edu/\~{}ccallan/TCRPaper/genes/}.


\section{Pre-selection model}
The pre-selection, or generative model, assumes the following structure for the probability distribution of recombination scenarios $S$ \cite{Murugan02102012}:
\beq\label{eq:pgen}
\begin{split}
P_{\rm pre}(S)=&P(V)P(D,J)P({\rm insVD})P({\rm insDJ})\\
&\ P({\rm delV}|V)P({\rm dellD},{\rm delrD}|D)P({\rm delJ}|J)\\
&\ P(s_1)P(s_2|s_1)\cdots P(s_{\rm insVD}|s_{\rm insVD-1})\\
&\ P(t_1)P(t_2|t_1)\cdots P(t_{\rm insDJ}|t_{\rm insDJ-1}),
\end{split}
\eeq
where a scenario is given by the VDJ choice, the number of insertions insVD, insDJ and the number of deletions (delV,dellD), (delrD,delJ) at each of the two junctions, together with the identities $(s_1,\ldots,s_{\rm insVD})$,$(t_1,\ldots,t_{\rm insDJ})$ of the inserted nucleotides. It is worth noting that the insertions are assumed to be independent of the identities of the genes between which insertions are made. By contrast, the deletion probabilities are allowed to depend on the identity of the gene being deleted. These validity of these assumptions is verified {\sl a posteriori}.

\section{Model fitting}
\subsection{Maximum likelihood formulation}
The model probability to observe a given coding nucleotide sequence is:
\beq
P_{\rm post}(\vec{\tau},V, J)={Q(\vec{\tau},V, J)}{P_{\rm pre}(\vec{\tau},V, J)},
\eeq
where $\vec{\tau}=(\tau_1,\ldots,\tau_{3L})$ is the nucleotide sequence of the CDR3 (defined as running from the conserved cysteine in the V segment up to the last amino acid in the read, leaving two amino acids between the last read amino acid and the conserved phenylalanine in the J segment), $L$ is the length of the CDR3, and $V$ and $J$ index the choice of the germline $V$ and $J$ segments (which completely determine the sequence outside the CDR3 region). The $D$ segment is entirely absorved into $\vec\tau$, and is not explicitly tracked in assessing selection.
The selection factor $Q$ is assumed to take the following factorized form:
\beq
{Q(\vec{\tau},V, J)}=\frac{1}{Z} \ q_L \   q_{V,J} \ \prod_{i=1}^L q_{i;L}(a_i).
\eeq
where $\vec{a}=(a_1,\ldots,a_L)$ is the amino-acid sequence of the CDR3, and $Z$ is a normalization constant that enforces
\beq
\sum_{\vec{\tau},V,J}P_{\rm post}(\vec\tau,V,J)=1.
\eeq

The probability, $P_{\rm pre}(\vec \tau,V,J)$, of generating a specific sequence in a V(D)J recombination event can be obtained from the noncoding sequence reads by the methods explained in \cite{Murugan02102012}. Specifically, the pre-selection model gives the probability $P_{\rm pre}(S)$ of a recombination scenario $S=(V,D,J,{\rm insVD},{\rm insDJ},{\rm delV},\ldots)$ as given by Eq.\,\ref{eq:pgen}. A scenario $S$ completely determines the sequence $\vec \tau$, but the converse is not true. The pre-selection probability for a coding sequence is thus given by
\beq
P_{\rm pre}(\vec \tau,V,J)=\frac{1}{p_{\rm coding}}\sum_{S\to (\vec \tau, V,J)} P_{\rm pre}(S)
\eeq
where we sum over scenarios resulting in a particular CDR3 sequence $\vec \tau$ and a particular $V,J$ pair. The normalization factor $p_{\rm coding}\approx 0.26$ corrects for the fact that a randomly generated sequence is not always productive ({\em i.e.} in-frame and with no stop codon). From this point on, we regard the initial generation probability of any specific read as known. When we make statements about the pre-selection distribution of CDR3 properties, such as length or amino acid utilization, they are derived from synthetic repertoires drawn from the above pre-selection distribution.

\begin{figure}
\ifthenelse{\equal{\onlycaption}{true}}{}{
\noindent\includegraphics[width=.7\linewidth]{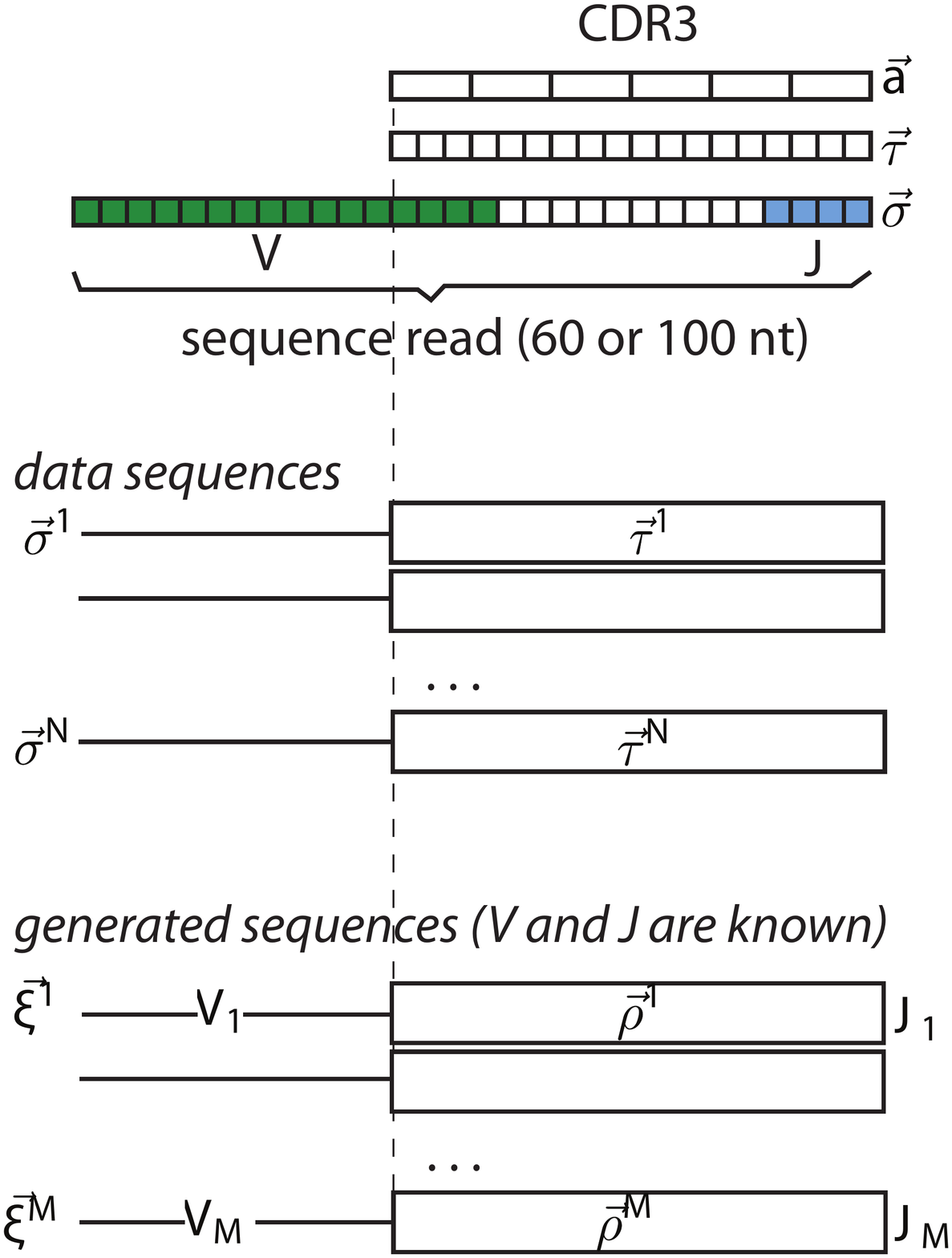}
}
\caption{Summary of the notations used in this paper for the sequences. The CDR3 region is defined from the conserved cysteine around the end of the V segment to the last amino-acid in the read, leaving two amino acids to the conserved phenylalanine in the J segment. The nucleotides in the read are defined as $\sigma_i$, the nucleotides in the CDR3 region as $\tau_i$ and the amino acids in the CDR3 region as $a_i$. The data sequences therefore can be defined in terms of $\vec{\sigma}$, or  their $V$, $J$ genes and $\vec{\tau}$. The generated sequences, with known $V$ and $J$ genes, are defined in terms of $\vec{\xi}$ for the whole sequence or $\vec{\rho}$ for only the CDR3.
\label{fig:notations}
}
\end{figure}

We want to infer the parameters $q_L$, $q_{V,J}$ and $q_{i;L}(\cdot)$ of the model from the observed coding sequence repertoires. Formally we want to maximize the likelihood of the data given the model. Unfortunately the sequence reads from the data are not long
enough to fully specify the V and J segments, so we cannot use $P_{\rm post}(\vec\tau,V,J)$ as our raw likelihood. Instead, we need to write the probability of observing a given (truncated) read $\vec\sigma$, of length 60 or 101 nucleotides, depending on the donor:
\beq
P_{\rm post}(\vec\sigma)=\sum_{(V,J,\vec\tau)\to \vec\sigma} P_{\rm post}(\vec \tau, V,J).
\eeq
where we note again that $(\vec \tau, V,J)$ fully specifies $\vec\sigma$, while $\vec\sigma$ fully specifies $\vec\tau$, but not V and J.
Given a dataset of $N$ sequences, $\vec\sigma^1,\ldots,\vec\sigma^N$
(see Fig.~\ref{fig:notations} for notations), the likelihood reads:
\beq
\mathcal{L}(Q)=\prod_{a=1}^N P_{\rm post}({\vec\sigma}^a).
\eeq
Our goal is maximize $\mathcal{L}$ with respect to the parameters  $q_L$, $q_{V,J}$, and $q_{i;L}(\cdot)$ (globally refered to as $Q$).

\subsection{Expectation maximization}
Calculating $P_{\rm post}({\vec\sigma})$ is computationally intensive. Given the form of the model, it seems more natural to work with $P_{\rm post}(\vec \tau, V,J)$, but this likelihood involves the ``hidden'' variables $V$ and $J$.
To circumvent this problem, we use the expectation maximization
algorithm \cite{Dempster:1977ul,McLachlan:2008wo}. This algorithm uses
an iterative two-step process, with two sets of model parameters $Q$
and $Q'$. The log-likelihood of the
data is calculated using
the set of parameters $Q'$; in the ``Expectation'' step, this
log-likelihood is averaged over
the hidden variables with their posterior probabilities, which are
calculated using the second set of parameters $Q$. In the ``Maximization'' step, this average
log-likelihood is maximized over the first set $Q'$, while keeping the second
set $Q$ fixed. Then $Q$ is updated to the optimal value of $Q'$, and
the two steps are repeated iteratively until convergence.

In practice, starting with a test set of parameters $Q$, we calculate, for each sequence of the data, the posterior probability of a $(V,J)$ pair:
\beq
P_{\rm post}(V_a,J_a|\vec\sigma^a)=\frac{Q(\vec\tau^a,V_a,J_a)P_{\rm pre}(\vec\tau^a,V_a,J_a)}{\sum_{V,J} Q(\vec\tau^a,V,J)P_{\rm pre}(\vec\tau^a,V,J)}.
\eeq
The log-likelihood, expressed in terms of the hidden variables $V$ and $J$, is maximized after averaging over $V$ and $J$ using that posterior. Specifically we will maximize:
\beq
\begin{split}
&\mathcal{\hat L}(Q'|Q)=\sum_{a=1}^N\left\langle \log P_{\rm post}(\vec \tau^a, V_a,J_a;Q')\right\rangle_{Q}\\
&\ \equiv\sum_{a=1}^N \sum_{V^a,J^a}P_{\rm post}(V_a,J_a|\vec\sigma^a;Q) \log P_{\rm post}(\vec \tau^a, V_a,J_a;Q').
\end{split}
\eeq
Here we have added the $Q$ dependencies explicitly because there are two different parameter sets $Q$ and $Q'$.
The maximization is performed over $Q'$, which parametrizes the log-likelihood itself, while keeping $Q$, which parametrizes how the average is done over the hidden variables, constant. After each maximization step we substitute:
\beq
Q\leftarrow \mathrm{argmax}_{Q'}\mathcal{\hat L}(Q'|Q),
\eeq
and iterate until convergence. This procedure is guaranteed to find a local maximum of the likelihood $\mathcal{L}(Q)$.

\begin{figure*}
\noindent\includegraphics[width=\linewidth]{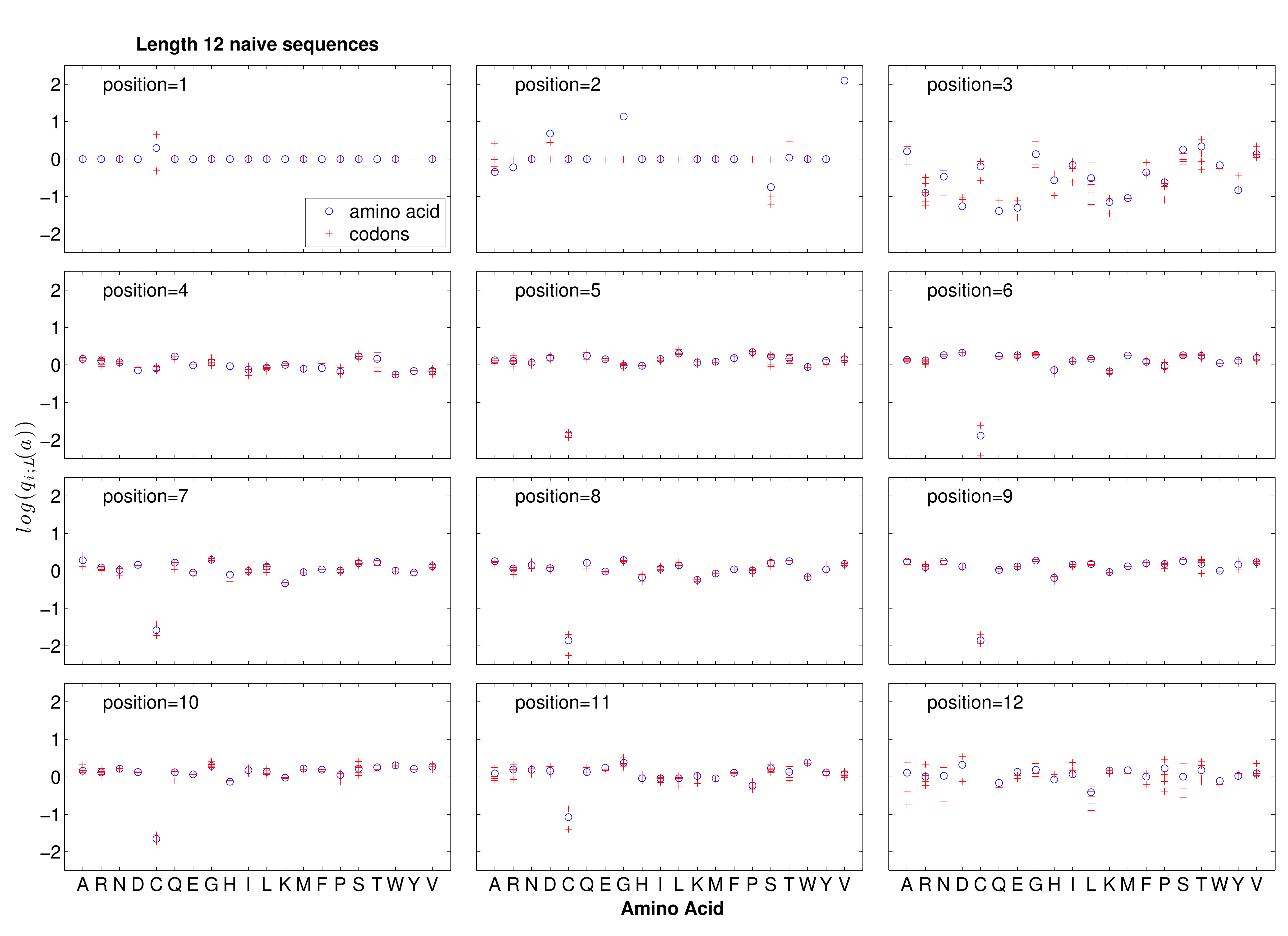}
\caption{The $q_{i;L}(a)$ selection factors learned for codons (red crosses) agree with those learned for amino acids (blue). The $q_{i;L}(a)$ are plotted for each position in the CDR3 region (panels from $1$ to $12$) for naive CDR3 sequences of length $12$, as a function of the amino acids at each position. A given amino acid at a given position can come from different codons, which are marked by multiple crosses at that position. Codons or amino acids for which there was not enough data to infer the selection factors are not represented.
\label{fig:codons}
}
\end{figure*}

\begin{figure*}
\noindent\includegraphics[width=.25\linewidth]{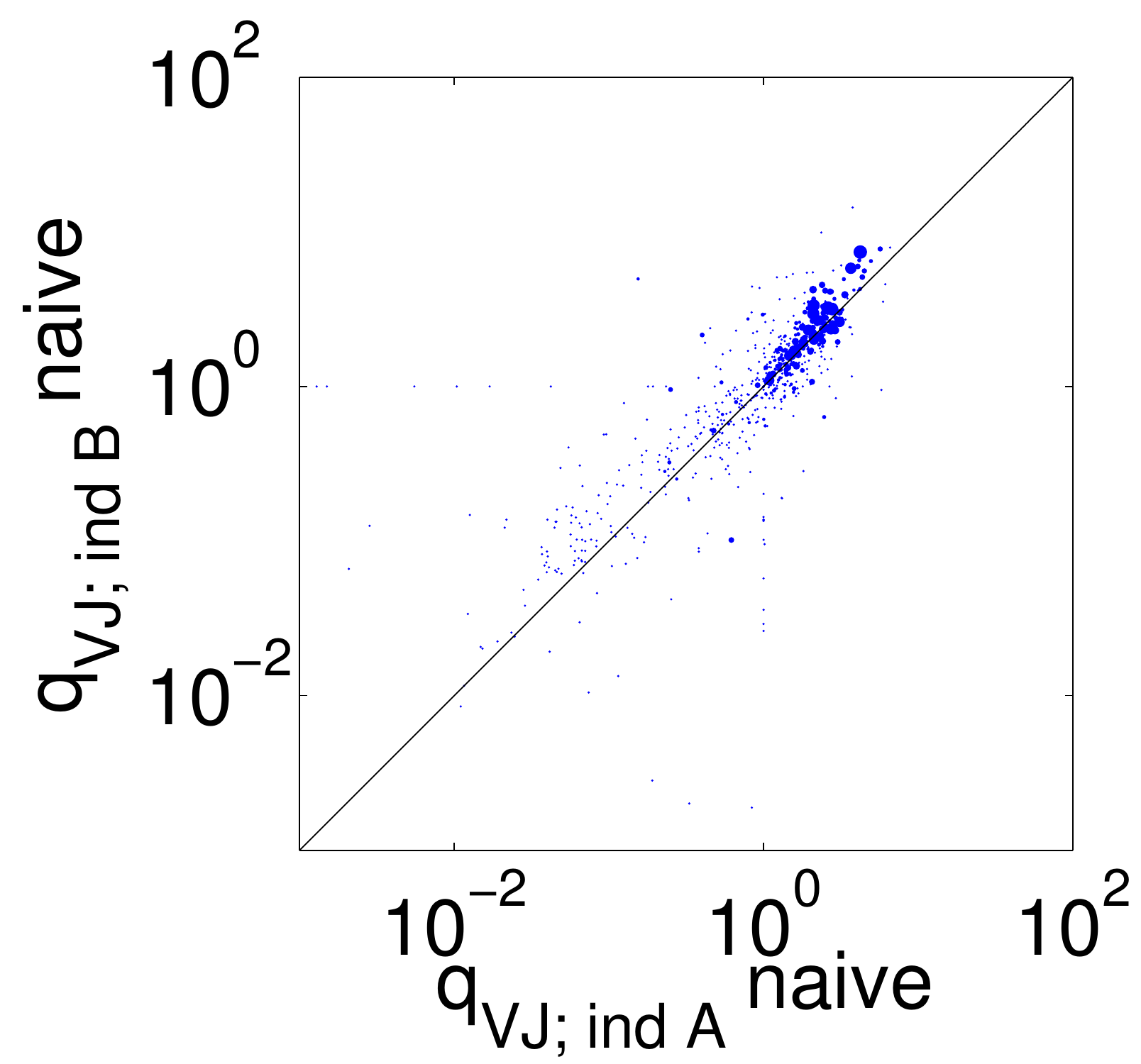}
\noindent\includegraphics[width=.25\linewidth]{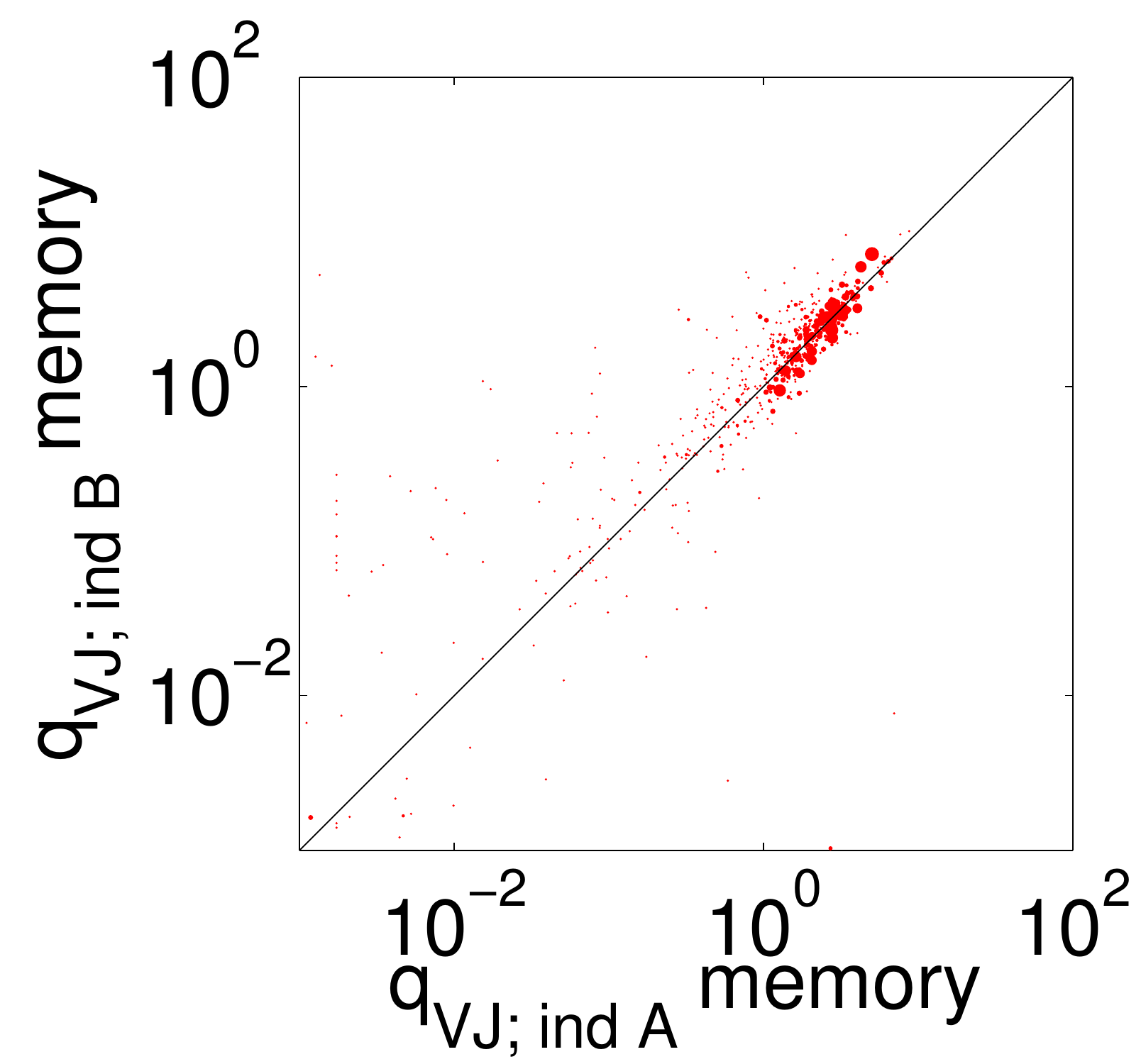}
\noindent\includegraphics[width=.25\linewidth]{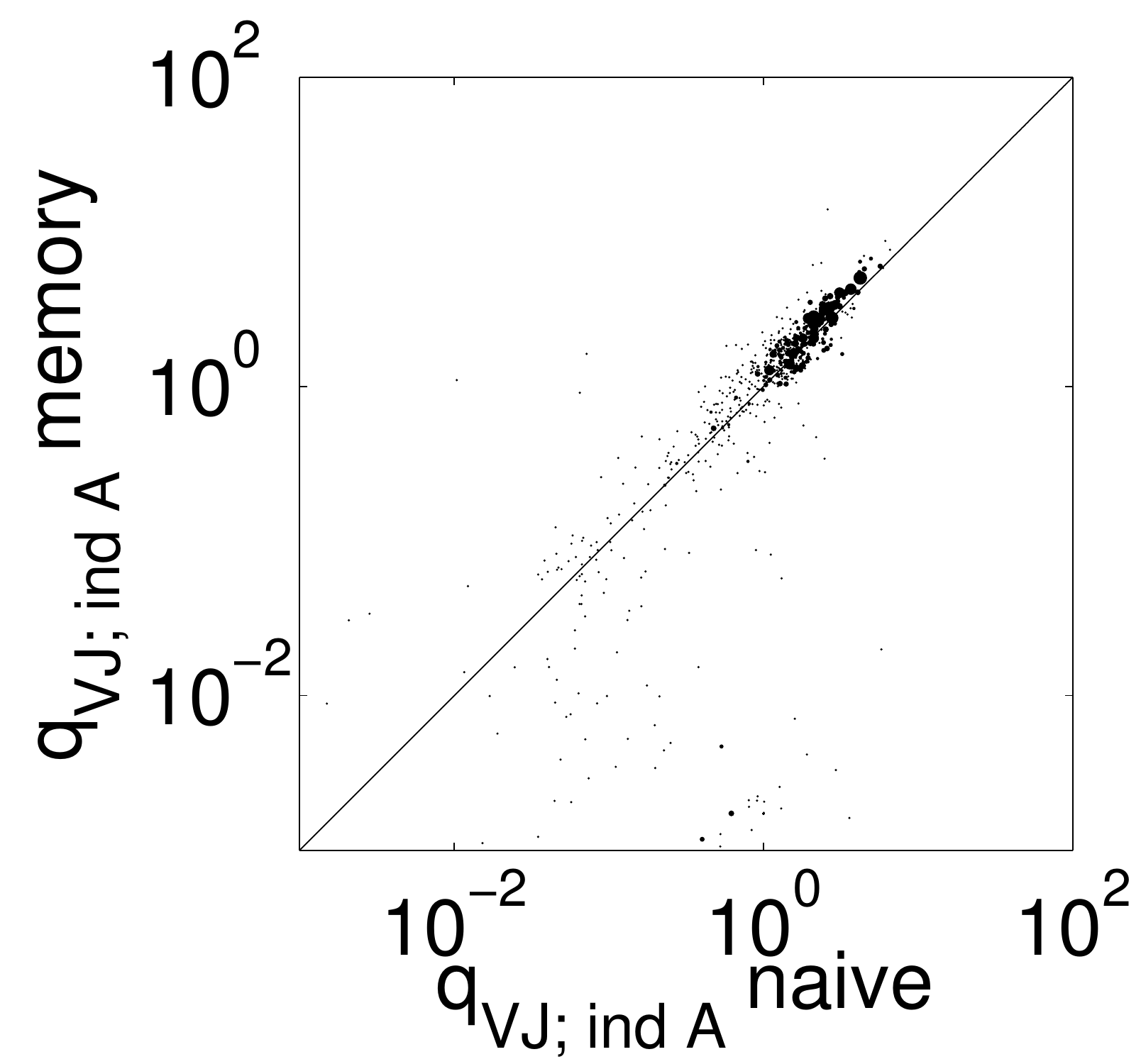}
\caption{The scatter of VJ gene selection factors $q_{VJ}$ between donors $A$ and $B$ for naive ({\bf A}) and memory repertoires ({\bf B}), as well as between the memory and naive repertoires of the same individual ({\bf C}) shows that the memory and naive repertoires are statistically similar to each other and across individuals. See Fig.~\ref{fig:corr} 
 for the correlation analysis of all individuals and cell types.
\label{fig:compqVJ}
}
\end{figure*}

\begin{figure}
\noindent\includegraphics[width=\linewidth]{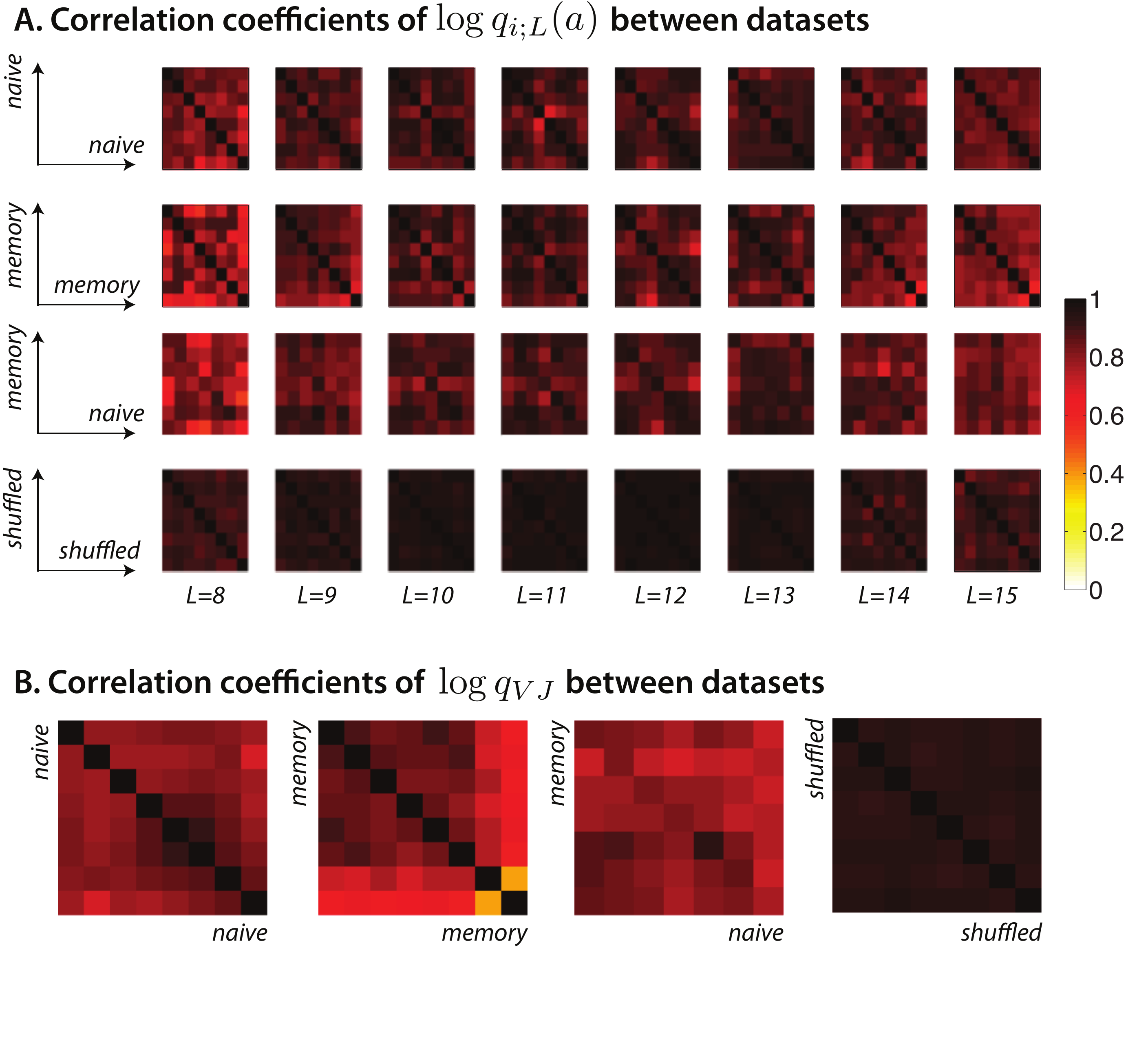}
\caption{Correlation coefficients between selection factors obtained for models learned for different donors and cell type (naive and memory). The compared factors are the amino-acid selection factors $q_{i;L}$ ({\bf A}) and the VJ gene selection factors $q_{VJ}$ ({\bf B}). Each position along the two axes in each plot corresponds to a different individual. The naive dataset of donor 8, and the memory dataset of donor 5 were removed because of too low statistics. In all heat maps, the $x$ and $y$ axes correspond to different donors (1-7;9 for naive, 1-4;6-9 for memory, and 1,2,3,4,6,7,9 for comparison between naive and memory).
\label{fig:corr}
}
\end{figure}

\subsection{Equivalence with fitting marginal probabilities}
The expectation-maximization step can be simplified by noting that at the maximum, derivatives vanish:
\beq
\frac{\partial\mathcal{\hat L}(Q'|Q)}{\partial Q'}=0.
\eeq
Precisely, we take derivatives with each of the parameters, $q_L$,
$q_{VJ}$ {\em etc.} and set them to zero. Since $P_{\rm post}(\vec
\tau, V,J)$ is naturally factorized in the $Q$ parameters, we obtain
simple expressions, {\em e.g.} $\partial \hat{\cal L}/\partial \log
q'_{L}=0$ gives:
\beq
\sum_{a=1}^N \sum_{V^a,J^a}P_{\rm post}(V_a,J_a|\vec\sigma^a;Q)\left(\delta_{L_a,L}-\frac{\partial \log Z}{\partial \log q'_{L}}\right)=0,
\eeq
where $\delta_{a,b}$ is Kronecker's delta function. The term in the
sum gives the total number of sequences in the data with length $L$. Besides
we have:
\beq
\frac{\partial \log Z}{\partial \log q'_{L}}=\sum_{\vec{\tau},V,J}\delta_{L(\vec\tau),L}P_{\rm post}(\vec\tau,V,J;Q')=P_{\rm post}(L;Q').
\eeq
Hence the maximality condition simply becomes:
\beq\label{eq:matchL}
P_{\rm data}(L)=P_{\rm post}(L;Q'),
\eeq
{\em i.e.} that the length distribution of the model must be equal to that of the data.
Similarly, maximizing with respect to $q_{i;L}(a_i)$ entails that single amino-acid frequencies at a given position are matched between data and model:
\beq\label{eq:matcha}
P_{i;L,{\rm data}}(a_i)=P_{i;L,\rm post}(a_i;Q').
\eeq
The condition for $q_{VJ}$ is slightly different, because we do not directly have the frequencies of $V$ and $J$ in the data. This is replaced by their expected frequency under the posterior $P_{\rm post}(V_a,J_a|\vec\sigma^a)$ taken with parameters $Q$:
\beq\label{eq:matchVJ}
\frac{1}{N}\sum_{a=1}^N P_{\rm post}(V,J|\vec\sigma^a;Q) = P_{\rm post}(V,J;Q'),
\eeq
where again the left-hand side is the empirical distribution of $V$ and $J$ (indirectly estimated with the help of the model with parameters $Q$), and the right-hand side is the model distribution of the same quantities (estimated with parameters $Q'$, which are then varied to achieve equality with the data estimate). The approach of iteratively adjusting model parameters to match a corresponding set of data marginals is a conceptually clear and computationally effective implementation of the expectation maximization algorithm.

\subsection{Gauge}
As defined above, the model is degenerate: for each $i,L$, the factors $q_{i;L}(a)$ and $Z$ may be multiplied by a common constant without affecting the model. We need to fix a convention, or gauge, to lift this degeneracy. We impose that, for each $i,L$:
\beq
\sum_{a=1}^{20} P_{i;L,\rm pre}(a)q_{i;L}(a)=1.
\eeq
where $P_{i;L,\rm pre}(a)$ is the probability of having amino-acid $a$ at position $i$ in CDR3s of length $L$.

\subsection{Numerical implementation}
To solve the fitting equations (\ref{eq:matchL})-(\ref{eq:matchVJ}) in practice, we use a gradient descent algorithm:
\beq
q_L\leftarrow q_L + \epsilon \left[P_{\rm data}(L)-P_{\rm post}(L;Q')\right],
\eeq
and similarly for $q_{i;L}$ and $q_{VJ}$. To do this, we must be able
to calculate the marginals $P_{\rm post}(L;Q')$, $P_{i;L,\rm
  post}(a_i;Q')$ and $P_{\rm post}(V,J;Q')$ from the model at each
step.

This leaves us with the problem of estimating marginals in the model, which we do using importance sampling.
Although it is easy to sample sequences from $P_{\rm pre}$ by picking
a random recombination scenario, sampling from $P_{\rm post}=QP_{\rm
  pre}$ is much harder, as the $q_{i;L}$, $q_L$ and $q_{VJ}$ factors
introduce complex dependencies between the different features of the
recombination scenario. To overcome this issue, we sample a large
number $M$ of $(\vec\tau,V,J)$ triplets from $P_{\rm
  pre}(\vec\tau,V,J)$, and, when estimating $P_{\rm post}$ expectation
values, weight the contribution of each sequence with its
$Q(\vec\tau,V,J)$ value (this is a particularly simple instance of
importance sampling). 
The generated triplets are denoted by
$[(\vec\rho^1,V_1,J_1),\ldots,(\vec\rho^M,V_M,J_M)]$, and the
corresponding reads by $(\vec\xi^1,\ldots,\vec\xi^M)$ (see
Fig.~\ref{fig:notations} for notations).
The marginal probability distribution of lengths, for instance, is estimated by
\beq\label{eq:modmarg}
P_{\rm post}(L;Q')\approx \frac{\sum_{b=1}^M \delta_{L_b,L} Q'(\vec\rho^b,V_b,J_b)}
{\sum_{b=1}^M Q'(\vec\rho^b,V_b,J_b)}.
\eeq
and similar expressions give estimates of $P_{i;L,\rm post}(a_i;Q')$ and
$P_{\rm post}(V,J;Q')$. Since we are optimizing over $Q'$,
the  sequences $(\vec\rho^b,V_b,J_b)$ can be generated once and
for all at the beginning of the algorithm. Then the marginal
probabilities are updated according to the modified $Q'$ using Eq.\,\ref{eq:modmarg}.
Finally, the normalization constant is evaluated by calculating:
\beq
Z\approx \frac{1}{M} \sum_{b=1}^M q_{L_b}q_{V_bJ_b} \prod_{i=1}^{L_b} q_{i;L_b}(a^b_i).
\eeq
so that 
\beq
\sum_{\vec\tau,V,J}P_{\rm post}(\vec\tau,V,J) \approx \frac{1}{M}
\sum_{b=1}^M Q(\vec\rho^b,V_b,J_b)=1.
\eeq

\subsection{Equivalence with minimum discriminatory information}
The principle of minimum discriminatory information is to look for a
distribution that reproduces exactly some mean observables of the
data, such as position-dependent amino-acid frequencies, while being
minimally biased with respect to some background distribution. When
the background distribution is uniform, this principle is equivalent
to the principle of maximum entropy.

Taking $P_{\rm pre}$ as our background distribution, assume we are looking for the
distribution $P_{\rm post}$ that satisfies
Eqs.~(\ref{eq:matchL})-(\ref{eq:matchVJ}) while minimizing
the divergence or relative entropy with respect to $P_{\rm pre}$, defined as:
\beq
D_{\rm KL}(P_{\rm post}\Vert P_{\rm pre})= \sum_{\vec\tau,V,J}
P_{\rm post}(\vec\tau,V,J)\log\frac{P_{\rm post}(\vec\tau,V,J)}{P_{\rm pre}(\vec\tau,V,J)}.
\eeq

Solving this problem is mathematically equivalent to solving the
maximum likelihood problem described above.

\begin{figure}
\noindent\includegraphics[width=.49\linewidth]{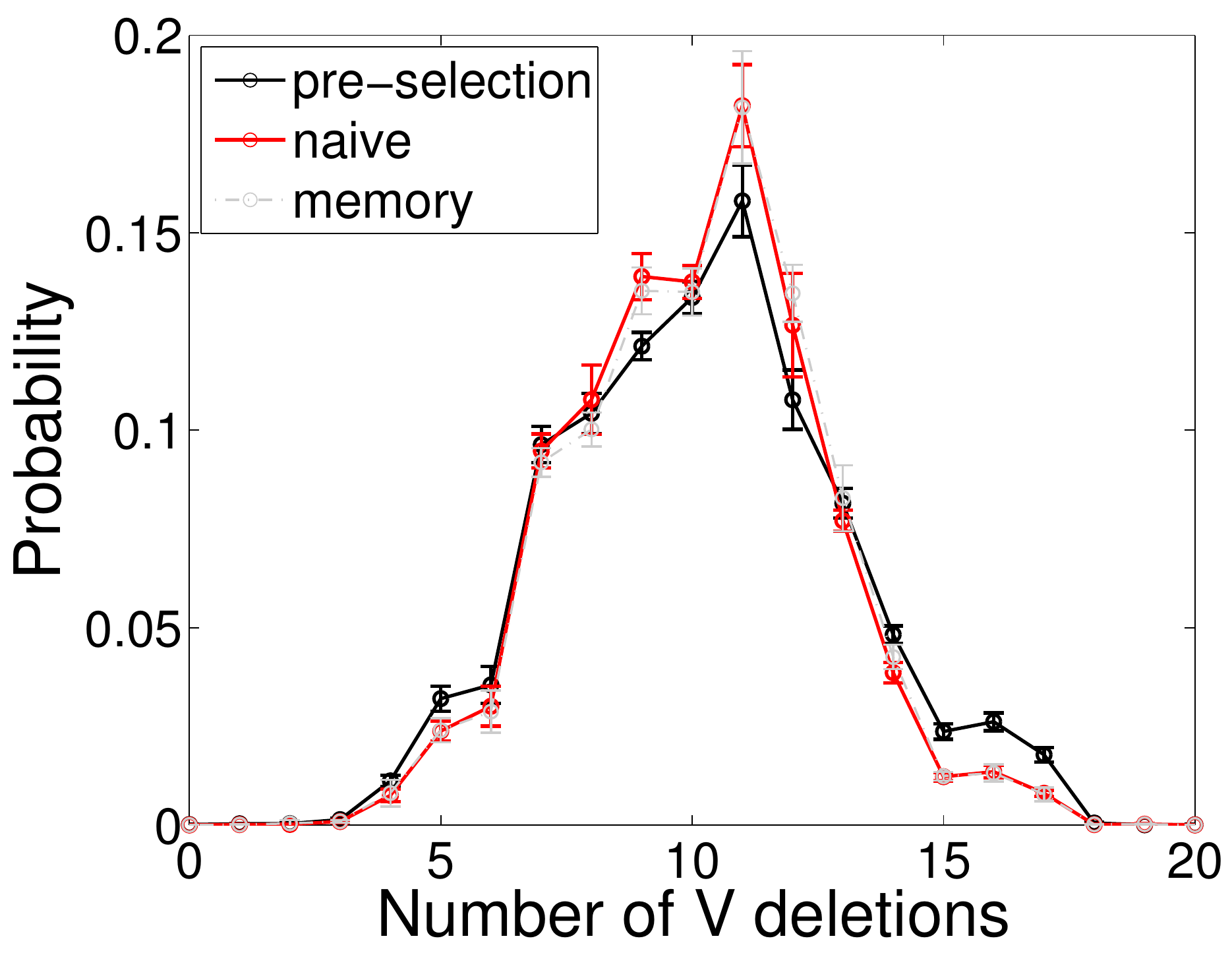}
\noindent\includegraphics[width=.49\linewidth]{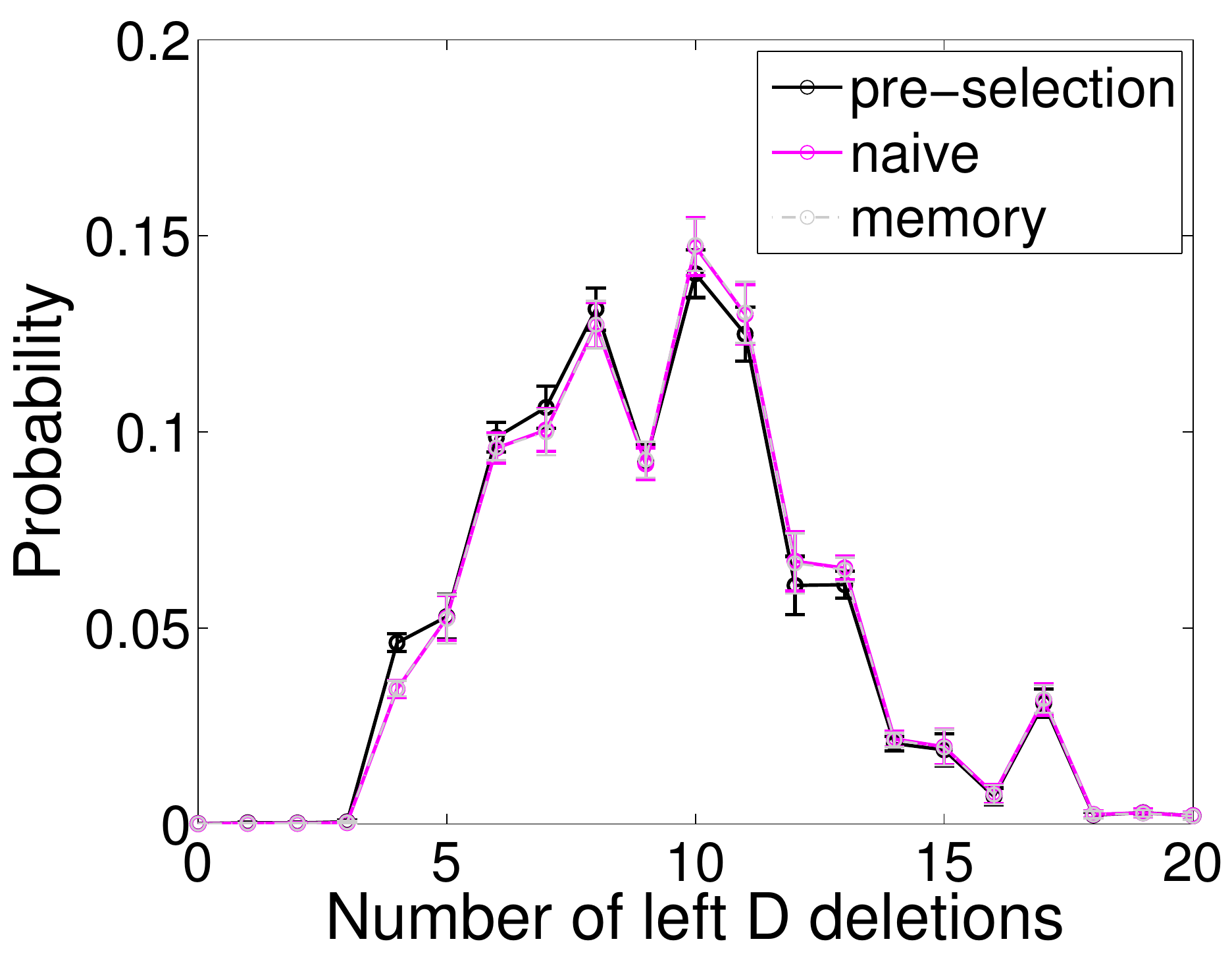}
\noindent\includegraphics[width=.49\linewidth]{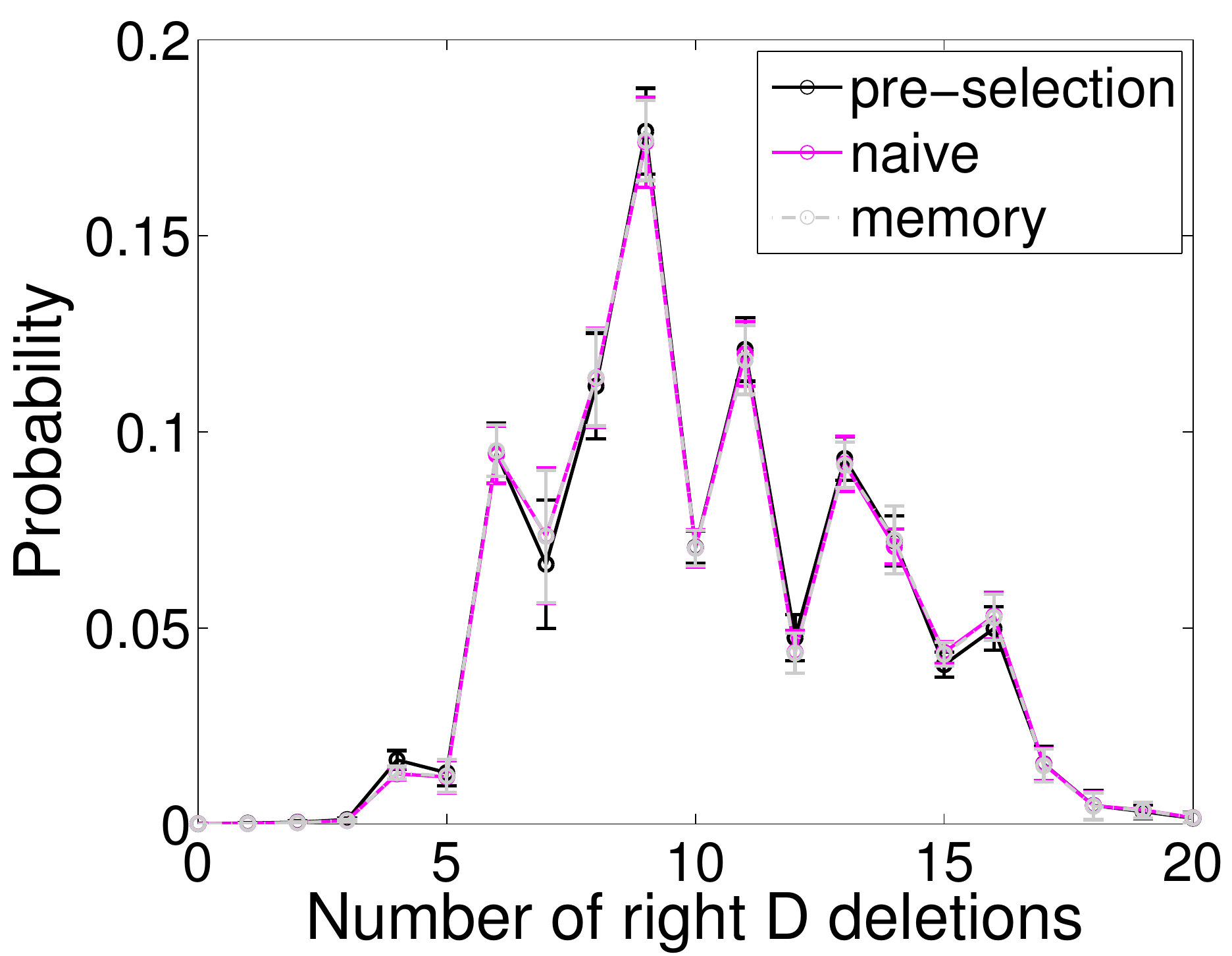}
\noindent\includegraphics[width=.49\linewidth]{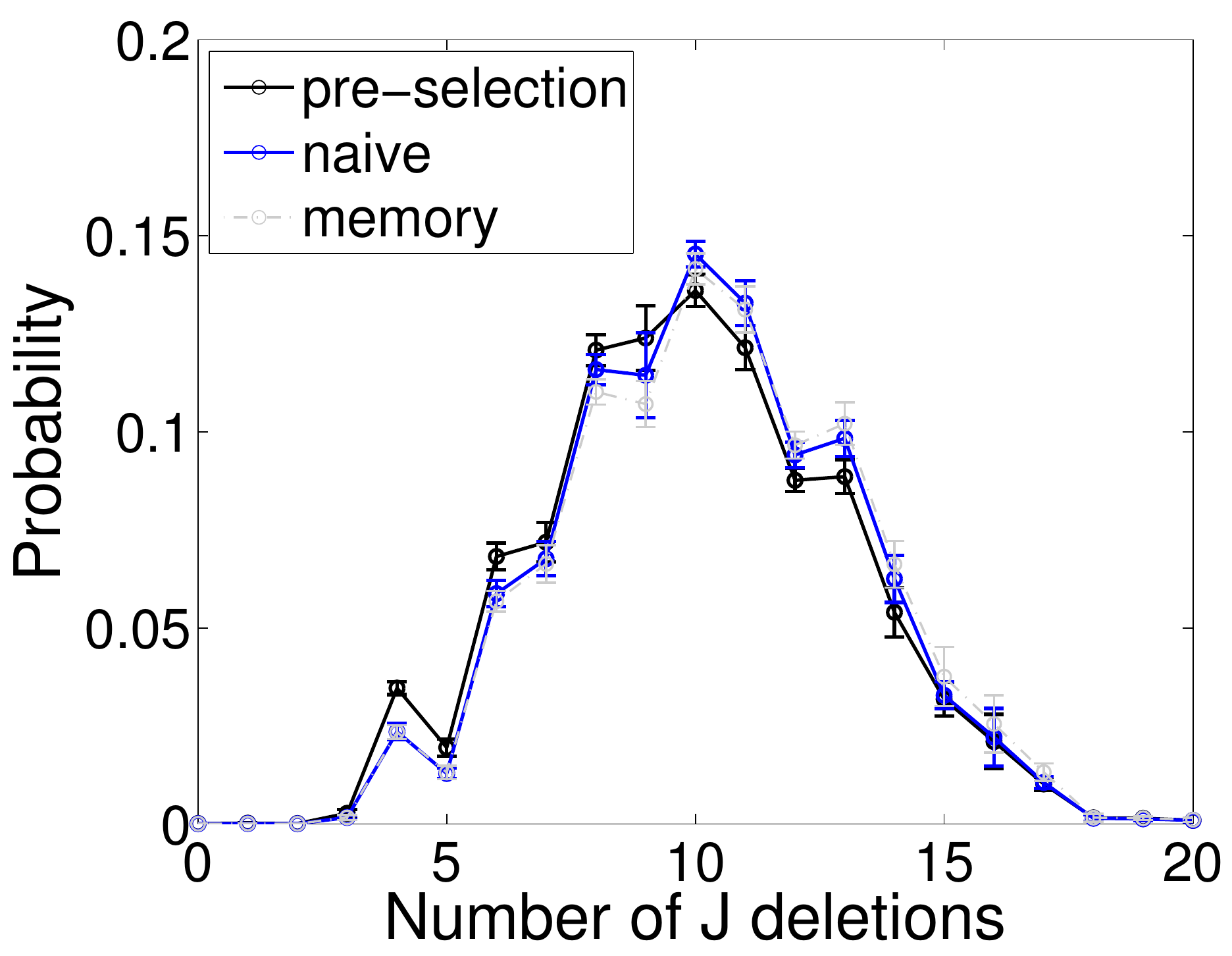}
\caption{The effects of selection on deletion profiles. Distribution of $V$ ({\bf A}), $D$ left-hand side ({\bf B}), $D$ right-hand side ({\bf C}), and $J$ ({\bf D}) deletions in the pre-selected (black lin
e), naive (colored line) and memory (gray dashed line) repertoires. Error bars show standard deviation over $9$ individuals. Results using $9$ separate models learned for each of the individuals. The deletion distributions for the memory repertoire are the same as for the naive repertoire. Selection has a slight effect on favoring distributions with non-extreme deletion values of deletions for $V$ and $J$ deletions, and does not have a significant effect on $D$ deletions.
\label{fig:del}
}
\end{figure}

\section{Individual, universal and shuffled donors}
We partition the data in three different ways to learn the
model. First, we learn a distinct model for each donor, and for each
of the naive and memory pools. For each donor, we have a distinct
$P_{\rm pre}$ learned from the out-of-frame sequences of that donor (although in fact they differ little from donor to donor as
discussed in \cite{Murugan02102012}). Second, we pool all the sequences of a
given type (naive or memory) from all nine donors together, and learn
a ``universal'' or average model. For this we use a mean $P_{\rm pre}$
averaged over all nine donors, and then learn $Q$ using all sequences. Third, to assess the effect of finite-size sampling in the universal model, we partition the data from all donors into nine random subsamples of equal sizes. This way we can estimate how much variability one should expect from just sampling noise.

\section{Entropy, distributions of $P_{\rm pre}$, $P_{\rm post}$ and $Q$}
To estimate global statistics, such as entropy, from the model, we draw a large set of sequences $(\vec\xi^1,\ldots,\ldots,\vec\xi^M)$ from $P_{\rm pre}$, and weight them according to the inferred (normalized) $Q$ values. Specifically, for each generated sequence, we estimate its primitive generation probability by summing over all the possible scenarios that could have given rise to it:
\beq
P_{\rm pre}(\vec\xi^b)=\frac{1}{p_{\rm coding}}\sum_{S\to \xi^b} P_{\rm pre}(S)
\eeq
where $\vec\xi^b$ is the full nucleotide sequence, including the CDR3 $\vec\rho^b$ as well as the $V_b$ and $J_b$ segments.
The entropy (in bits) of the selected sequence repertoire is defined as
\beq
H[P_{\rm post}]=-\sum_{\vec\sigma} P_{\rm post}(\vec\sigma)\log_2 P_{\rm
  post}(\vec\sigma)
\eeq
and, to include selection effects, we estimate it by
\beq
H[P_{\rm post}] \approx -\frac1M \sum_{b=1}^M Q(\vec\rho^b,V_b,J_b) \log\left[
Q(\vec\rho^b,V_b,J_b) P_{\rm pre}(\vec\xi^b)\right].
\eeq

The distributions of $P_{\rm pre}$, $P_{\rm post}$ and $Q$ over the selected sequences are
determined from the same draw of M sequences from $P_{pre}$, weighted by the normalized selection factors $Q$. For example
the distribution of $\log P_{\rm pre}$ is:
\beq
\mathbb{P}(\log P_{\rm pre})\approx \frac1M \sum_{b=1}^M
Q(\vec\rho^b,V_b,J_b) \delta\left[
  \log P_{\rm pre} - \log P_{\rm pre}(\vec\xi^b)\right].
\eeq

Marginal distributions over pairs of amino-acids $(a_i,a_j)$ at two positions $i$ and $j$
can also be calculated using the $\vec\rho^b$ sequences and weighting
them with $Q$. This can be generalized to arbitrary marginals or statistics.

\begin{figure}
\noindent\includegraphics[width=\linewidth]{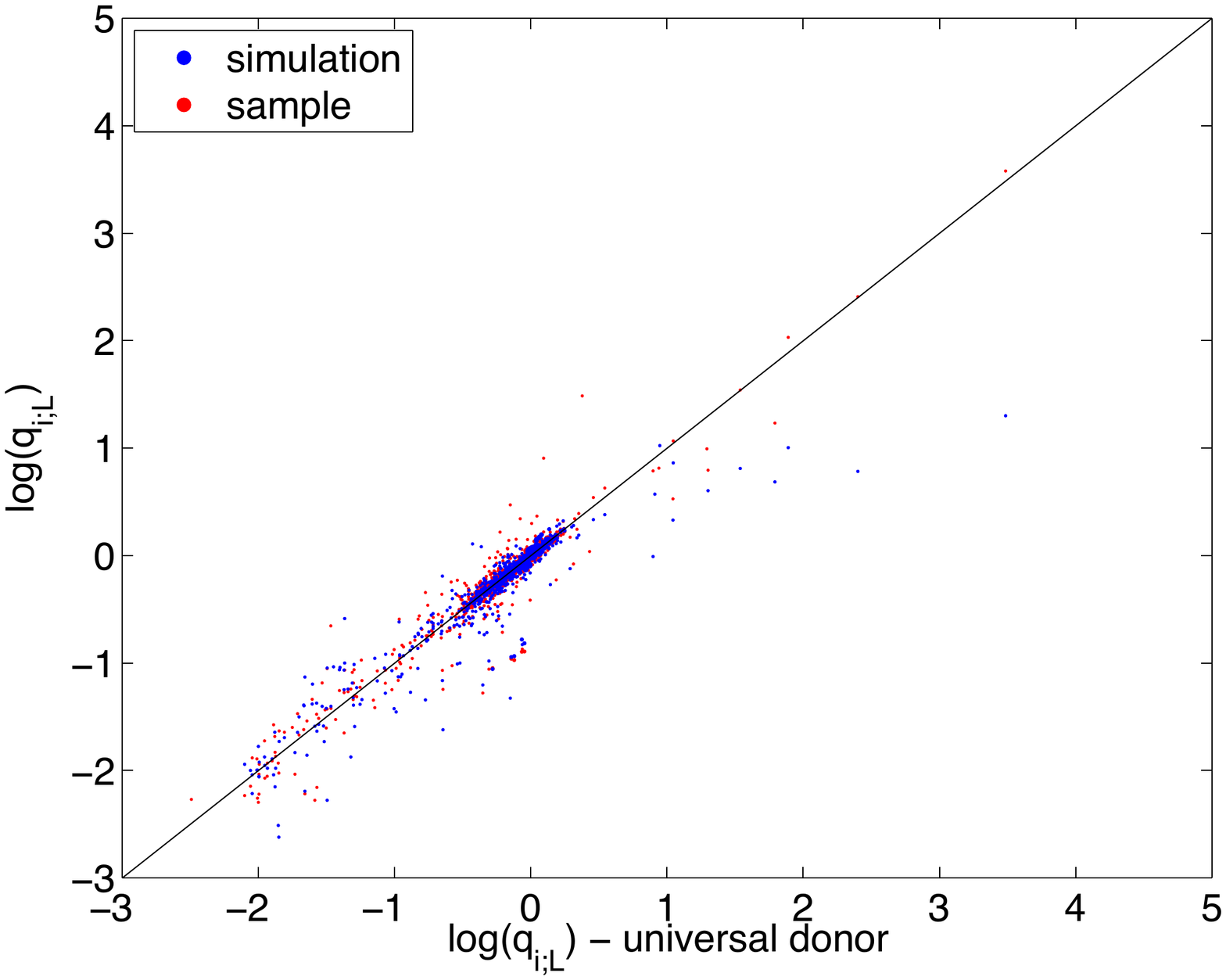}
\caption{The saturation of the $P_{data}(Q)/P_{pre}(Q)$ ratio does not affect the inference of the model. We simulated a dataset from $P_{\rm pre}$ and selected sequences with probability $\min[Q(\vec\sigma)/7,1]$. The plot compares the $q_{i;L}(a)$ selection factors directly inferred from data (ordinate) to values inferred from such simulated data (blue dots: simulation). The scatter in these points is compared to the scatter obtained from learning the selection factors using a random subset of the data (red dots: sample). The size of the points denotes the probability $P_{i;l,{\rm data}}(a)$ in the data repertoire. 
\label{fig:simul}
}
\end{figure}

\begin{figure*}
\noindent\includegraphics[width=.8\linewidth]{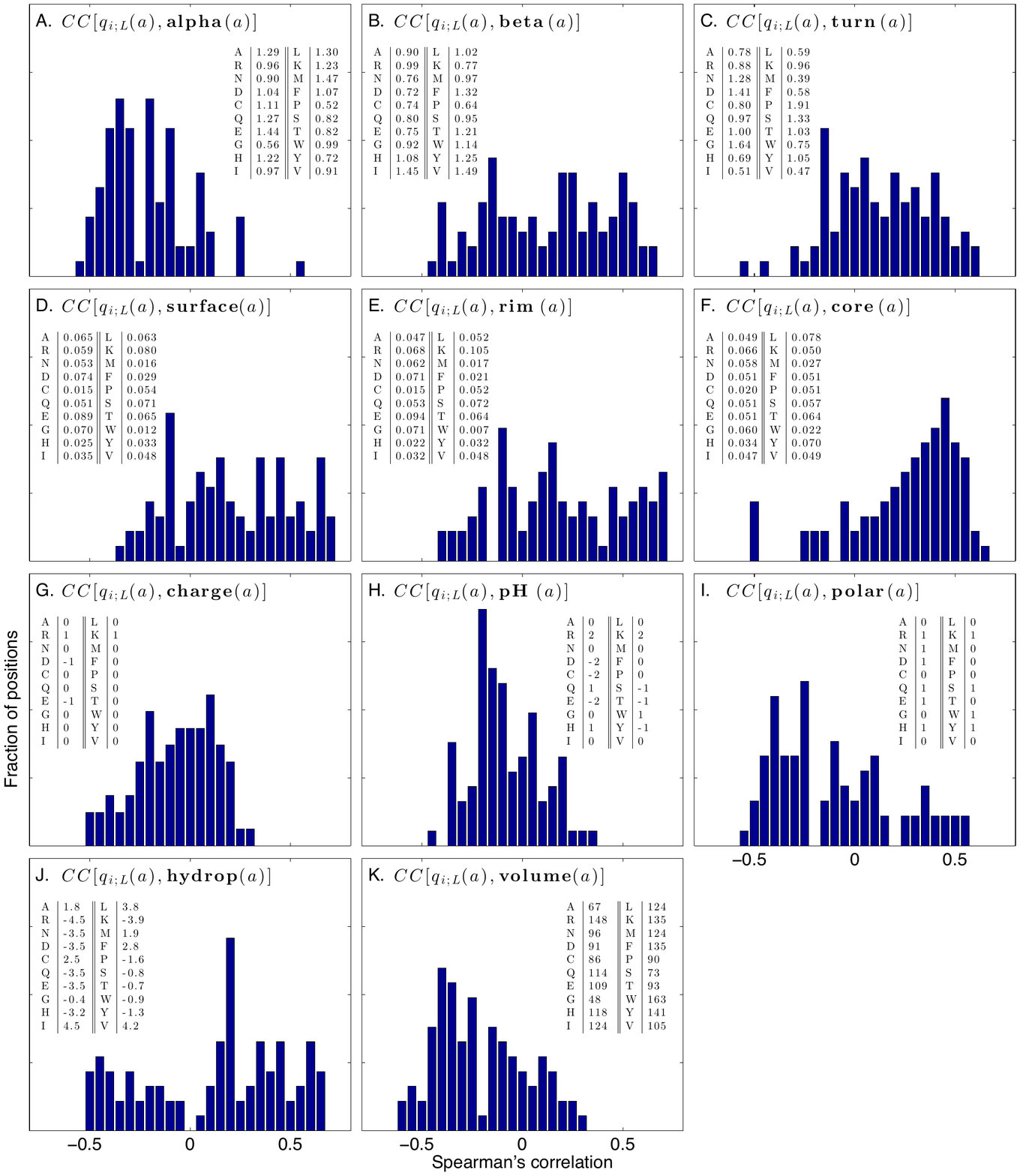}
\caption{Correlation of the $q_{i;L}$ selection factors with several biochemical properties. Each panel shows the histogram, over all positions and lengths, of Spearman's correlation coefficient between the $q_{i;L}(a)$ values for a given amino acid and the biochemical properties of that amino acid. The following biochemical properties are considered (from left to right, top to bottom): preference to appear in alpha helices ({\bf A}), beta sheets ({\bf B}), turns ({\bf C}) (source for ({\bf A-C}): Table 3.3 \cite{Stryer}). Residues that are exposed to solvent in protein-protein complexes (following definitions and data from \cite{MartinLavery}) are divided intothree groups: surface (interface) residues that have unchanged accessibility area when the interaction partner is present ({\bf D}), rim (interface) residues that have changed accessibility area, but no atoms with zero accessibility in the complex ({\bf E}) and core (interface) residues that have changed accessibility area and at least one atom with zero accessibility in the complex ({\bf F}). Rim residues roughly correspond to the periphery of the interface region, and core residues correspond to the center. Finally we plot the basic biochemical amino acid properties (source: {\tt http://en.wikipedia.org/wiki/Amino\_acid} and {\tt  http://en.wikipedia.org/wiki/Proteinogenic\_amino\_acid}): charge ({\bf G}), pH ({\bf H}), polarity ({\bf I}), hydrophobicity ({\bf J}) and volume ({\bf K}). For all properties the actual numerical values used to calculate the correlations are listed in the inset tables. We see a positive correlation trend with turns and core residues and a negative correlation trend with the preference of amino acids to appear in alpha helices and volume. 
\label{fig:biochem}
}
\end{figure*}

\section{Shared sequences}

The number of shared sequences in a subset of donors is counted based on the nucleotide sequences. This empirical number can then be compared to two kinds of theoretical predictions. Either by assuming that the sequences of each donor were generated and selected by a ``private'' model $P_{\rm post}^{(\alpha)}$, where $\alpha$ denotes the donor, {\em i.e.} a model inferred from the sequences of donor $\alpha$; or by assuming that sequences were generated and selected by a ``common'' or universal model $P_{\rm post}^{(u)}$ inferred from all sequences together. The latter is justified by the fact that differences between private models are small, and could reflect spurious noise that would exaggerate differences between individuals.

If we assume private models, the expected number of shared sequences between donors $\alpha$ and $\beta$ is:
\beq
N_{\alpha}N_{\beta}\sum_{\vec\sigma} P_{\rm post}^{(\alpha)}(\vec\sigma) P_{\rm post}^{(\beta)}(\vec\sigma),
\eeq
where $N_\alpha$ and $N_\beta$ are the numbers of sequences in each donor dataset.
To estimate that number, we
collect sequences that are shared between the generated datasets $\{\vec\xi^a\}$ of two (or more) donors, and reweight them by $Q$:
\beq
\frac{N_{\alpha}N_{\beta}}{M_\alpha M_{\beta}}\sum_{(\vec\rho,V,J)\in \alpha \cap \beta} Q^{(\alpha)}(\vec\rho,V,J)Q^{(\beta)}(\vec\rho,V,J),
\eeq
where $M_\alpha$ and $M_\beta$ are the number of generated sequences for each donor model, and where the sum is over the sequences found in the $\{\vec \xi^a\}$ dataset of both donors. Similar equations are used for comparing more than two donors.

If we assume a common model, the expected number of shared sequences reads:
\beq
N_{\alpha}N_{\beta}\sum_{\vec\sigma} [P_{\rm post}^{(u)}(\vec\sigma)]^2.
\eeq
This can be estimated by:
\beq
\frac{N_{\alpha}N_{\beta}}{M}\sum_{b=1}^M P_{\rm pre}^{(u)}(\vec\xi^b)[Q^{(u)}(\vec\rho^b,V_b,J_b)]^2,
\eeq
where $\{\vec\xi^a\}$ are sequences generated from the mean VDJ recombination model $P_{\rm pre}^{(u)}$. Similarly, the number of shared sequences between a triplet of donors $\alpha$, $\beta$, $\gamma$ is:
\beq
\frac{N_{\alpha}N_{\beta}N_\gamma}{M}\sum_{b=1}^M [P_{\rm pre}^{(u)}(\vec\xi^b)]^2[Q^{(u)}(\vec\rho^b,V_b,J_b)]^3,
\eeq
and likewise for quadruplets and more.

The expected numbers of shared sequences calculated above are averages. Their distribution is given by a Poisson distribution of the same mean. We use these Poisson distribution to estimate the error bars in Fig.~6A and \ref{fig:sharedmem}A, as well as the distributions in Fig.~6B-C and S\ref{fig:sharedmem}B-C.

\begin{figure}
\noindent\includegraphics[width=.49\linewidth]{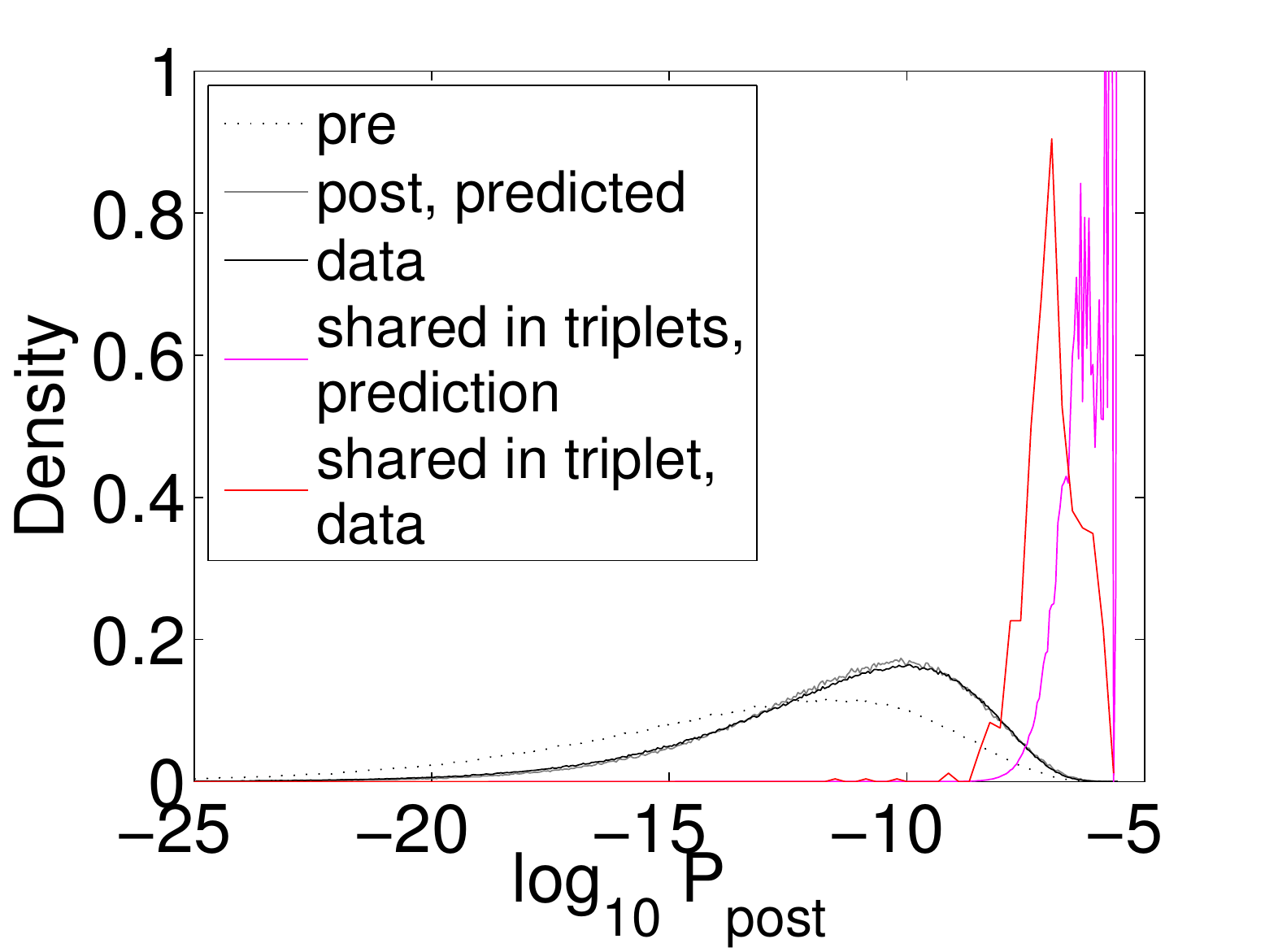}
\noindent\includegraphics[width=.49\linewidth]{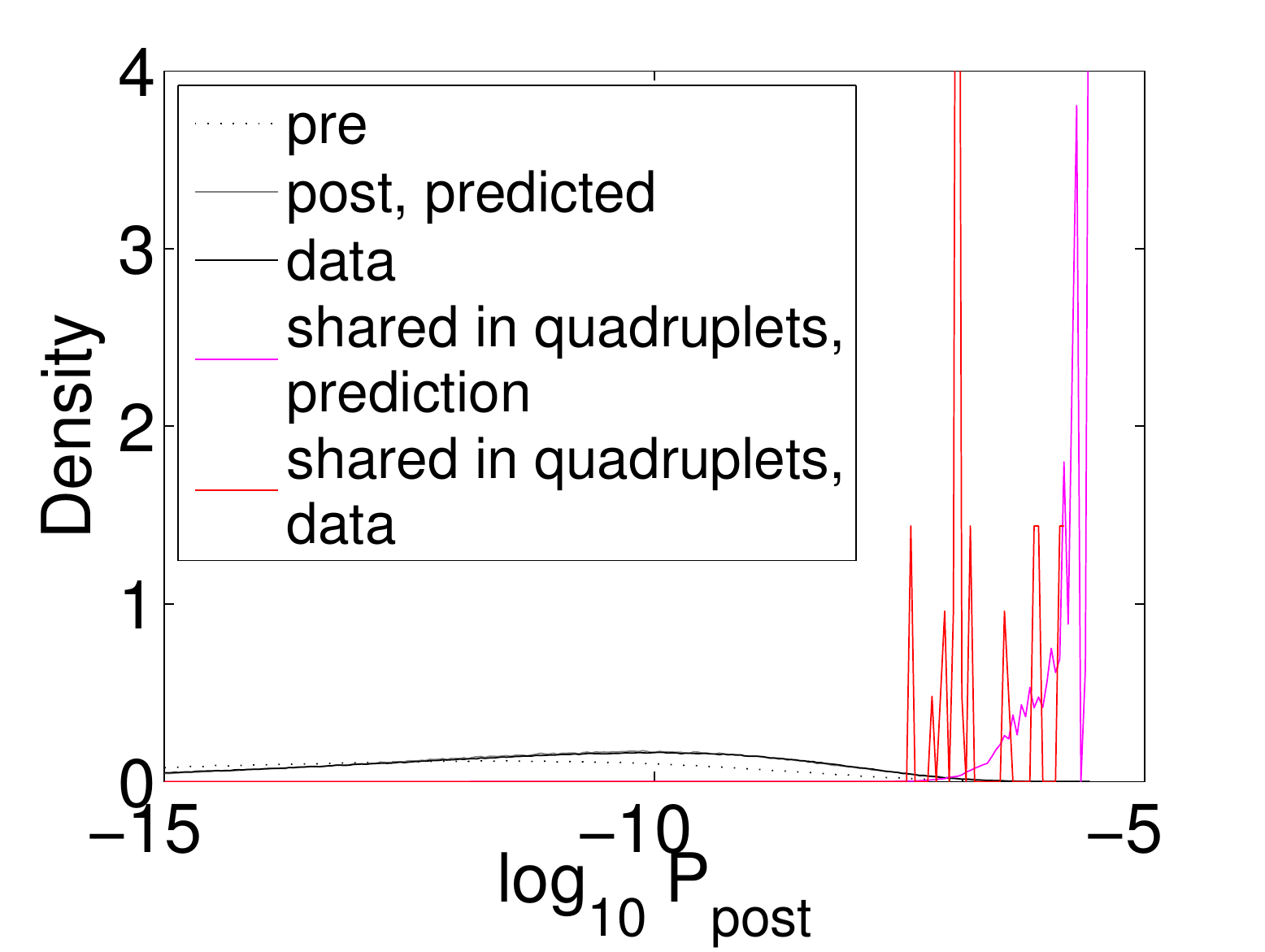}
\caption{Model prediction (magenta) and observed (red) distributions
of $P_{\rm post}$ in the naive sequences that are shared between at
least three (left) or four (right) donors. The model discrepancy may
be attributed to its failure to capture the very highly probable
sequences.
\label{fig:tripquad}
}
\end{figure}

\begin{figure}
\noindent\includegraphics[width=.99\linewidth]{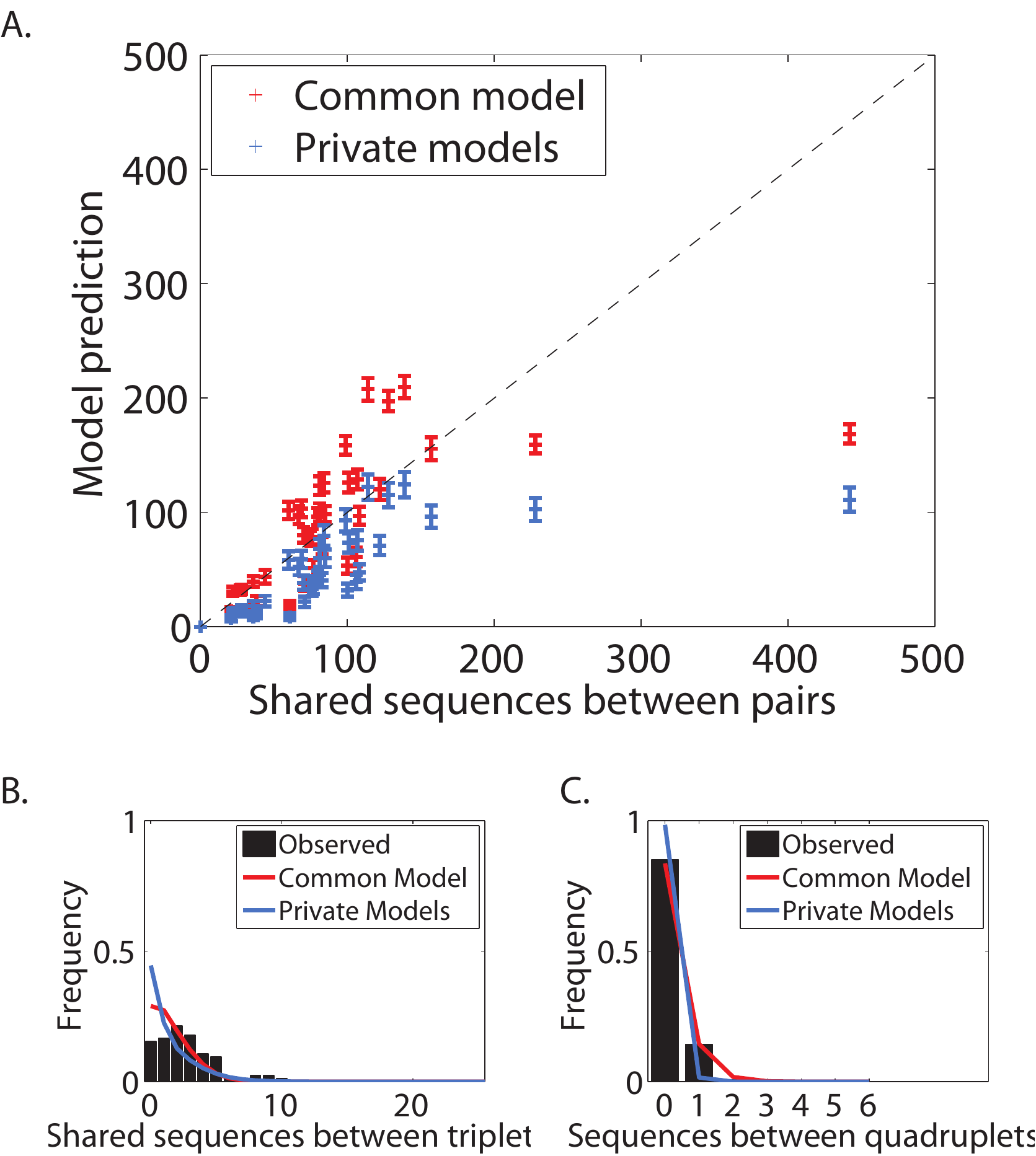}
\caption{Comparison between data and model for the number of shared sequences in the {\em memory} repertoires, in pairs ({\bf A}), triplets ({\bf B}) and quadruplets ({\bf C}) of individuals.
\label{fig:sharedmem}
}
\end{figure}

If we assume a common model, sequences that are shared between at least $n$ individuals are distributed according to $\propto [P^{(u)}_{\rm post}]^n$. To explore the statistics of these sequences, we take our $\vec\rho^b$ sequences generated from $P^{(u)}_{\rm pre}$ and weigh them with $[P^{(u)}_{\rm pre}(\vec\rho^b)]^{n-1}[Q^{(u)}(\vec\rho^b)]^n$. For example, to estimate the distribution of $\log P_{\rm post}$ in shared sequences as in Fig.~6D (for pairs), and Fig.~\ref{fig:tripquad} (for triplets and quadruplets), we calculate:
\beq
\begin{split}
\mathbb{P}(\log P_{\rm post})\approx &\frac1M \sum_{b=1}^M
[P^{(u)}_{\rm pre}(\vec\xi^b)]^{n-1}[Q^{(u)}({\vec\rho^b,V_b,J_b})]^n \\
&\times\delta\left[
  \log P_{\rm post} - \log P_{\rm post}^{(u)}(\vec\xi^b)\right].
\end{split}
\eeq

Sampling from shared sequences is equivalent to sampling from the high-probability, large deviation regime of the distribution. This statement can be made more physically intuitive by rewriting
$P_{\rm post}$ as a Boltzmann distribution $e^{-E/T}$ with $T=1$ and $E=- \log P_{\rm post}$. Considering sequences observed in at least $n$ donors, is equivalent to sampling from $(1/Z(n)) e^{-nE}$ (where $Z(n)$ is a normalisation constant), {\em i.e.} the Boltzmann distribution with $T=1/n$. Sequences shared between more and more individuals correspond to lower and lower temperatures, and thus lower energies and higher probabilities. In the low temperature regime, the roughness of the landscape depicted in Fig.~4C is starting to become important, and may not be well captured by our model, as suggested by Fig.\,\ref{fig:tripquad}.

\section{Codon model}
It is reasonable to assume that selection acts on the protein structure, at the amino acid level. But each amino acid can be obtained using a number of different codons, which could in principle each have a different selection factor.
We checked the robustness of our selection coefficients by learning an alternative model in which selection acts on codons. We present the results of this alternative codon model in Fig.~\ref{fig:codons} on the example of CDR3 sequences of length $12$. We show the  $q_{i;L}(a)$ selection factors at each position for each amino acid, and compare them to the selection factors obtained for the codons coding for that amino acid. We see that, especially in the bulk of the CDR3 sequence, selection at the level of codons or amino acids are equivalent, proving the generality of our approach.

\section{Additional effects of selection on repertoire properties}
In the main text we present several repertoire properties, such as insertion profiles and comparisons of the $q_{i;L}(a)$ selection factors between naive and memory repertoires. In Fig.~\ref{fig:del} we plot the deletion profiles for $V$, $J$ and $D$-lefthand side and $D$-righthand side deletions, comparing the distributions for the pre-selection, naive and memory repertoires. We note that the deletion profiles for the $V$ and $J$ distributions are more peaked, favoring intermediate deletion values. However the $D$ distributions are little affected by selection. Similarly to the case of insertion distributions shown in in Fig.~3E-F, the naive and memory distributions appear indistinguishable within the error bars.

In Fig.~3A-C, the selection factors $q_{i;L}(a)$ acting on amino acids are compared between individuals and cell type. Similarly, the selection factors acting on the genes $q_{VJ}$ are statistically indistinguishable between the memory and naive repertoires for one individual, compared to the variability between the naive (or memory) repertoires taken from two sample individuals (see Fig.~\ref{fig:compqVJ}).

To compare the repertoires of individuals as well as the naive and memory repertoires with each other, we consider the correlation coefficients between the selection factors $\log q_{i;L}$, and between the VJ gene selection factor $\log q_{VJ}$, of different individuals (Fig.~\ref{fig:corr}). Correlations between memory and naive repertoires are similar to those between naive-naive or memory-memory repertoires for different individuals; all are a bit smaller than the correlations between the artificial, shuffled sequence datasets, where the discrepancy is entirely attributable to statistical noise. These observations lead us to the conclusion that at this level of description, the selection processes that shape the memory and naive repertoires are very similar with each other and between different individuals.

\newpage

\section{Effects of saturation of global selection factors on the
  inference procedure}

We consider distributions of the selection factor $Q$ in the
pre-selection ensemble $P_{\rm pre}(Q)$, in the post-selection ensemble according to
the model $P_{\rm post}(Q)$, and in the actual data sequences $P_{\rm data}(Q)$. These three distributions are formally defined as:
\beqn
P_{\rm pre}(Q)&=& \frac{1}{M}\sum_{b=1}^M
\delta\left[
 Q - Q(\vec\rho^b,V_b,J_b)\right].\\
P_{\rm post}(Q)&=& \frac{1}{M}\sum_{b=1}^M Q(\vec\rho^b,V_b,J_b)
\delta\left[
  Q - Q(\vec\rho^b,V_b,J_b)\right] \\
&= &Q P_{\rm pre}(Q).\\
P_{\rm data}(Q)&=& \frac{1}{N}\sum_{a=1}^N\sum_{V_a,J_a}P_{\rm post}(V_a,J_a|\vec\sigma^a)\nonumber\\
&&\times\delta\left[
 Q - Q(\vec\tau^a,V_a,J_a)\right]
\eeqn

As can be seen in Fig.~4, the ratio of the
distribution of global selection factors  $P_{\rm data}(Q)/P_{\rm pre}(Q)$ saturates for large values of $Q$. To make sure that this saturation does not impair our ability to correctly infer the selection factors, we simulated a dataset from $P_{\rm pre}$ and selected sequences with probability $\min[Q(\vec\sigma)/7,1]$ to mimic the effects of this plateau. We then inferred the selection coefficients for this artificial dataset. We see that the saturation does not affect our ability to correctly infer the selection coefficients (Fig.~\ref{fig:simul}) and the variability in the inferred $q_{i;L}(a)$ selection factors is of the same order as from using random subsamples of the original data.

\section{Biochemical correlations}
To check for correlations of our inferred $q_{i;L}(a)$ selection factors with known biochemical properties, we calculated Spearman's coefficient between the selection factors and a number of standard quantities (see Fig.\,\ref{fig:biochem} for the full list). We find that the selection factors do not correlate well with most standard properties, such as charge, hydrophobicity and polarity. However we do find a trend of positive correlation with amino acids that are likely to appear in turns (Fig.\,\ref{fig:biochem} C) and ones that have been identified as those that make the core of the interface in a protein-protein complexes (Fig.\,\ref{fig:biochem} F) \cite{MartinLavery}. We find a trend of negative correlations with amino acids that have large volume (Fig.\,\ref{fig:biochem} K) and are likely to appear in alpha helices (Fig.\,\ref{fig:biochem} A). These observations are consistent with the fact that structurally CDR3 regions form loops and bulky amino acids as well as stabilizing alpha helix-like interactions would interfere with this structure. Core amino acids are at the center of the interface and are known to be the main contributors to interface recognition and  affinity. On the other hand interface rim and non-interface (surface) residues, which  are both in touch to various degrees with the solvent and are not crucial interface forming elements, show similar non-distinctive correlation patterns.

\bibliographystyle{pnas}
\bibliography{biblio,frompapers,VDJ_paper}

\ifthenelse{\equal{\format}{pnasfigend}}{\panelone}{}
\ifthenelse{\equal{\format}{pnasfigend}}{\paneltwo}{}
\ifthenelse{\equal{\format}{pnasfigend}}{\panelthree}{}
\ifthenelse{\equal{\format}{pnasfigend}}{\panelfour}{}
\ifthenelse{\equal{\format}{pnasfigend}}{\panelfive}{}
\ifthenelse{\equal{\format}{pnasfigend}}{\panelsix}{}

\end{document}